\pdfoutput=1
\documentclass[superscriptaddress,aps,prd,twocolumn,nofootinbib,notitlepage]{revtex4-1}

\usepackage{graphics,graphicx,amsthm,amsmath,amssymb}
\usepackage{color}
\usepackage{mathtools}
\usepackage{hyperref}
\input{Qcircuit}
\usepackage{tikz}
\usepackage[]{algorithm2e}
\usepackage[procnames]{listings}

\definecolor{keywords}{RGB}{255,0,90}
\definecolor{comments}{RGB}{0,0,113}
\definecolor{red}{RGB}{160,0,0}
\definecolor{green}{RGB}{0,150,0}
 
\def\({\left(}
\def\){\right)}

\def\bra#1{\mathinner{\langle{#1}|}}
\def\ket#1{\mathinner{|{#1}\rangle}}

\def\CZ{{\rm CZ}}

\def\T{{\rm T}}
\def\X{{\rm X}}
\def\Y{{\rm Y}}

\def\H{{\rm H}}

\def\mE{{\mathbb E}}

\def\exp{{\rm exp}}

\def\cU{{\mathcal U}}

\newcommand{\isep}{\mathrel{{.}\,{.}}\nobreak}

\begin{document}

\title{Simulation of low-depth quantum circuits as complex undirected graphical
  models}

\author{Sergio Boixo}
\affiliation{Google Inc., Venice, CA 90291, USA}
\author{Sergei V. Isakov}
\affiliation{Google Inc., Venice, CA 90291, USA}
\author{Vadim N. Smelyanskiy}
\affiliation{Google Inc., Venice, CA 90291, USA}
\author{Hartmut Neven}
\affiliation{Google Inc., Venice, CA 90291, USA}

\begin{abstract}
  Near term quantum computers with a high quantity (around 50) and
  quality (around 0.995 fidelity for two-qubit gates) of
  qubits will approximately sample from certain probability
  distributions beyond the capabilities of known classical algorithms
  on state-of-the-art computers, achieving the first milestone of
  so-called quantum supremacy. This has stimulated recent progress in
  classical algorithms to simulate quantum circuits. Classical
  simulations are also necessary to approximate the fidelity of
  multiqubit quantum computers using cross entropy benchmarking. Here
  we present numerical results of a novel classical simulation algorithm to
  calculate output probabilities of universal random circuits with more qubits and depth than
  previously reported. For example, circuits with $5 \times 9$ qubits of depth 40, $7 \times 8$ qubits of depth 30, and 
  $10 \times (\kappa > 10)$) qubits of depth 19 are all easy to sample
  by calculating around one thousand measurements in a single workstation. Cross entropy benchmarking with around
  one million measurements for these circuits is now also possible in
  a computer cluster. The algorithm is related to the ``Feynman
  path'' method to simulate quantum circuits. For low-depth circuits, the algorithm scales
  exponentially in the depth times the smaller lateral
  dimension, or the treewidth, as explained in
  Boixo et. al.~\cite{boixo2016characterizing}, and therefore confirms the
  bounds in that paper. In particular, circuits with $7 \times 7$
  qubits and depth 40 remain currently out of reach. Follow up work on
  a supercomputer environment will tighten this bound.
\end{abstract}

\date{\today}

\maketitle

\section{Introduction}
Quantum computers offer the promise to perform a number of
computational tasks believed to be impossible for classical
computers. Prominent examples are quantum
simulations~\cite{feynman1982simulating,aspuru2005simulated,hastings2015improving,wecker2015solving,o2016scalable,reiher2017elucidating,babbush2017low,jiang2017quantum}
and factoring large numbers~\cite{shor1999polynomial}. Given the progress in classical
computation, surpassing the capabilities of state-of-the-art
algorithms and computers is no small feat. Therefore, in the path to
more practical ``quantum supremacy'' breakthroughs, it is reasonable
to focus first on more specific milestones better suited for near-term
quantum computers with a high quantity (around 50) and quality (around
0.995 two qubit gate fidelity) of qubits. Sampling from the output
distribution of certain families of quantum circuits has been
identified as a promising venue for a first milestone of quantum
supremacy~\cite{aaronson2011computational,bremner2015average,bremner16,gao2017quantum,hangleiter2017direct,kapourniotis2017nonadaptive}. Universal random circuits~\cite{boixo2016characterizing} (or more general
chaotic quantum evolutions~\cite{neill2017}) are particularly well-suited for
near term devices. On the one hand, we note that an important
characteristic of these types of sampling problems is that they are
hypersensitive to perturbations~\cite{schack_hypersensitivity_1993,scott_hypersensitivity_2006} and, therefore, approximate
classical algorithms are unlikely to work. On the other hand, this
also implies that an experimental implementation will not be
successful unless the qubits being used are very coherent and all
operations have high fidelity (and very fine-tuned
control). Nevertheless, the same hypersensitivity to perturbations
makes this type of demonstration of quantum supremacy a good benchmark
for multiqubit fidelity of universal computers using cross entropy
benchmarking~\cite{boixo2016characterizing,neill2017}. A successful implementation of a first milestone
of quantum supremacy would demonstrate the basic building blocks for a
large-scale quantum computer within the operational range of the
surface code~\cite{barends_superconducting_2014,fowler2012surface}.

The prospect of a near-term demonstration of quantum supremacy has
also stimulated recent progress in classical algorithms to simulate
quantum circuits~\cite{de2007massively,boixo2016characterizing,zulehner2017advanced,haner2017petabyte,pednault2017breaking}. The most efficient algorithms previously
reported to simulate generic circuits (without symmetries that allow
for faster compressions~\cite{viamontes2009quantum,zulehner2017advanced} or ``emulations'') apply gates to a
vector state, implemented as highly optimized matrix-vector
multiplications~\cite{de2007massively,boixo2016characterizing,haner2017petabyte,pednault2017breaking}.
Benchmarking results for a highly optimized algorithm of
this type were reported in Ref.~\cite{boixo2016characterizing} for
$6 \times 7$ qubits and depth 27 taking 989 seconds and performed in
the Edison supercomputer. More recently, Ref.~\cite{haner2017petabyte}
reported a simulation using 8192 nodes and 0.5 petabytes of memory for
a quantum circuit (in the same ensemble) with $5 \times 9$ qubits and
depth 25 on the new Cori II supercomputer in 552 seconds (and $6
\times 7$ qubits in 80 seconds). Finally, Ref.~\cite{pednault2017breaking} recently reported simulations of
a circuit with $7 \times 7$ qubits and depth 27 and another circuit of
$7 \times 8$ qubits and depth 23, along two days in the IBM Blue Gene/Q
supercomputer.
Computing time in a supercomputer at that scale is scarce and
expensive.

We generalize the simulation of a quantum circuit by mapping it onto
an undirected graphical model, which we compute using the variable
elimination algorithm, developed in the context
of exact inference~\cite{dechter1998bucket,murphy_machine_2012,murphy_machine_2012}. In some cases a circuit amplitude can be mapped directly
to the partition function of an Ising
model with imaginary temperature~\cite{boixo2016characterizing}, which
can be computed exactly with the same method. This is related to the
Feynman path method~\cite{bernstein1997quantum}, the tensor network method~\cite{markov_simulating_2008}, and similar
approaches~\cite{aaronson2016complexity}. The algorithm is explained
in Sec.~\ref{sec:ve}, and we provide pseudocode in App.~\ref{app:pseudocode}. We are mostly interested in sampling output
probabilities, but the same method can be used to calculate many
amplitudes at once or calculating observables exactly. The
quantum circuits implemented with superconducting qubits, which is our
main focus, use controlled-phase gates (and controlled-Z gates in
particular) as the predominant two-qubit
gate~\cite{martinis_fast_2014,barends_superconducting_2014}. We
exploit the fact that controlled-phase gates are diagonal in the
computational basis (this is also used in the algorithms referred
above).  The cost of this algorithm is exponential in
$\min(O(d \ell), O(n))$ for depth $d$, minimum lateral dimension
$\ell$ and total number of qubits $n$. More optimally, the cost is
exponential in the treewidth, which is upper bounded by
$\min(O(d \ell), O(n))$, and was estimated in
Ref.~\cite{boixo2016characterizing} (see also
Sec.~\ref{sec:treewidth}).

\section{Numerical results}\label{sec:nums}

In the simplest implementation, we compute  one-by-one the exact probabilities assigned by a
quantum circuit $\cU$ to a chosen
set of outputs. One use case is cross entropy
benchmarking~\cite{boixo2016characterizing,neill2017}, where we need to
compute the probabilities assigned by $\cU$ to the bit-strings measured in an 
experiment, see App.~\ref{app:ceb}. In other use cases, such as estimating the value of an observable, or simulating
sampling from the output distribution, the error scales as $t^{-1/2}$,
where $t$ is the number of output probabilities computed (see
App.~\ref{app:mcerror}). If the time per
amplitude is small enough (for instance, around 100 seconds or less), then we
can obtain enough probabilities in a single workstation in a
reasonable time (for instance, 1000 probabilities in around one
day). The sampling is also trivially parallelizable across
machines. The algorithm can
be modified to sample sets of probabilities in a single run, see Eq.~\eqref{eq:psi_x}. It can also be modified to calculate an
observable exactly, but the computational cost might change. This
algorithm is not useful in all cases. For instance,
Ref.~\cite{boixo2016characterizing} numerically verified that the
output distribution of low-depth universal random circuits approaches
the entropy of the Porter-Thomas (or exponential) distribution to
within its standard deviation of order $2^{-n/2}$. It would require
to compute a number of output probabilities of order $2^n$ to verify
this property. We also note that
for enough depth a direct evolution of the wave vector is likely to
become more efficient, as it is easier to optimize. 

In what follows we report the time per probability (or amplitude) in a single
workstation. We use as a reference a machine with 2$\times$ 14-core
Intel E5-2690 V4 processors @ 2.6GHz, 35MB Cache and 128GB DDR4
2400MHz RAM. We use single precision complex numbers, as the relative error per
probability is negligible for the sizes computed (less then
$10^{-5}$). The computation of a single probability can also be
distributed among several machines as the size of the computation
grows. We leave this and other optimizations (such as reusing part of
the computation when calculating many probabilities) for follow up
work.

We are interested in circuits in a 2D lattice, as they are available
experimentally~\cite{martinis_fast_2014,barends_superconducting_2014}. We will
use the following set of gates:
 \begin{enumerate}
  \item We use controlled-phase (CZ) gates as two-qubit gates. 
   \item For a concrete example we use single qubit gates  in the set
     $\{\H,\X^{1/2}, \Y^{1/2}, \T\}$. The gate $\H$ is a Hadamard, $\X^{1/2}$ ($\Y^{1/2}$) is a $\pi/2$ rotation around the $\X$ ($\Y$) axis of the Bloch sphere, and the non-Clifford T gate is the diagonal matrix $\{1,e^{i \pi/4}\}$. 
  \end{enumerate}
We use the rules for the layout of gates described
in~\cite{boixo2016characterizing}. We are also particularly interested
in superconducting qubits, and it is currently not possible to perform two CZ gates simultaneously in two neighboring qubits~\cite{barends_superconducting_2014,barends_digital_2015,kelly_state_2015,barends_digitized_2015}. This restriction was used for the circuits.

\begin{figure}
  \centering
  \includegraphics[width=\columnwidth]{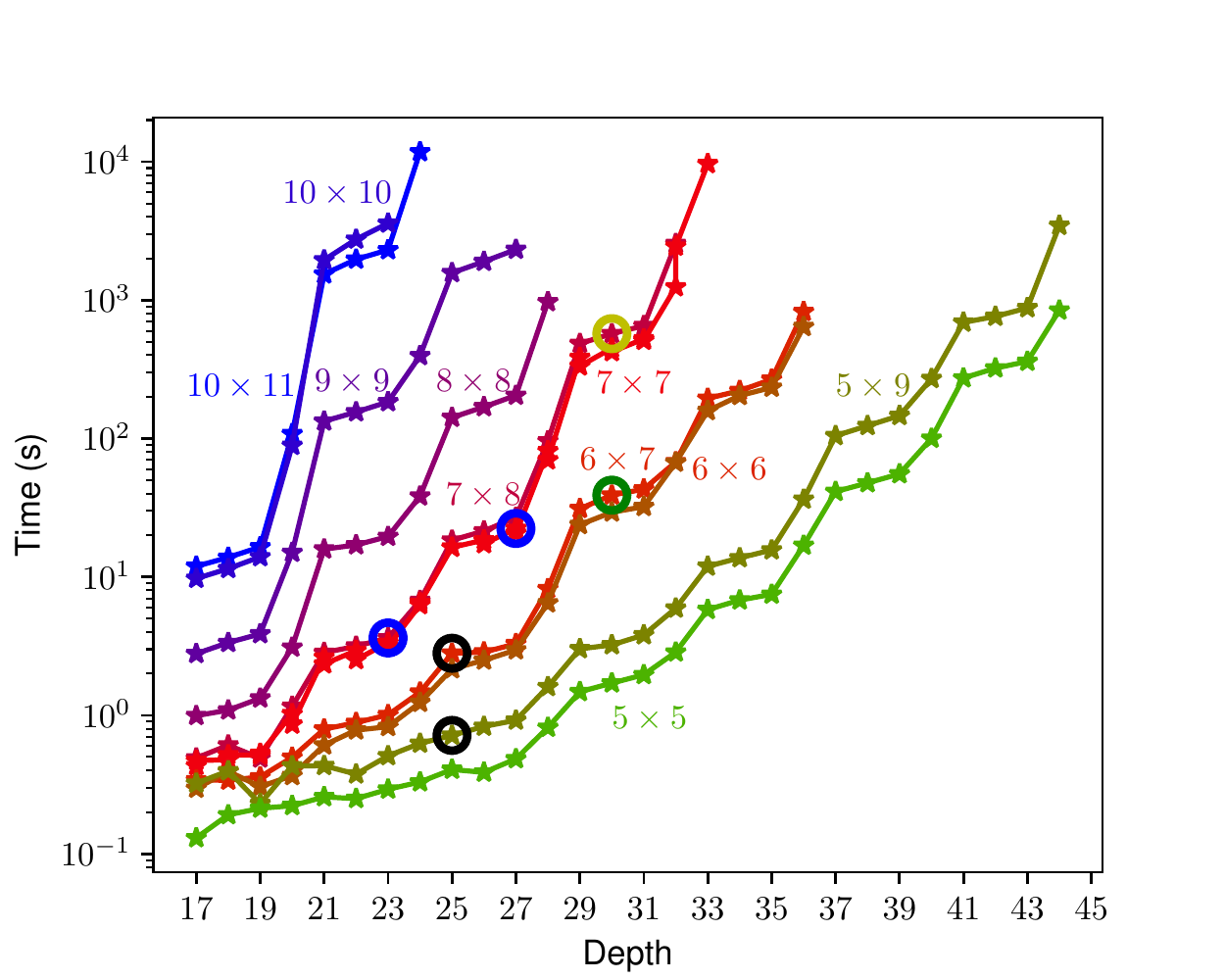}
  \caption{Time per output probability for a typical instance as
    a function of depth on a single workstation, using 
    TensorFlow as the engine for the computation and a vertical
    elimination ordering (see Sec.~\ref{sec:sveo}). Different colors
    corresponds to different circuit sizes, from $5 \times 5$ (and $5
    \times 9$) qubits, to $10 \times 10$ ($10 \times 11$). The green circle
    marker corresponds to the circuits simulated in
    Ref.~\cite{boixo2016characterizing}, black circles to
    Ref.~\cite{haner2017petabyte}, and blue circles to
    Ref.~\cite{pednault2017breaking}. The yellow circle is the circuit
  used in Fig.~\ref{fig:pt}.}
  \label{fig:timings_sizes}
\end{figure}

Figure~\ref{fig:timings_sizes} shows the time per output probability for typical instances of different circuit sizes as
    a function of depth on a single workstation.\footnote{To be
      precise, at depth 1 we apply the first layer of Hadamards for
      the circuits defined in Ref.~\cite{boixo2016characterizing}.} Different colors
    correspond to different circuit sizes, from $5 \times 5$ (and $5
   \times 9$) qubits, to $10 \times 10$ ($10 \times 11$). Here we use a
    vertical variable elimination ordering, which in essence processes
    the ``Feynman path'' or worldline of each qubit in sequence (see
    Sec.~\ref{sec:sveo}). The processing time (and required memory)
    of each computation grows exponentially with depth. The exponential
    growth is also faster for larger circuits. More specifically, we
    process the worldline of the qubits ordered first along the
    smaller lateral dimension $\ell$. The cost of
    computing probabilities for circuits with $5 \times 9$ qubits,
    with $\ell = 5$, is    similar to the cost of circuits with $5
    \times 5$ qubits. The same is true for circuits of sizes $7 \times
    7$ and $7 \times 8$ with $\ell=7$, and also sizes $10 \times 10$ and $10 \times 11$
    (or larger) with $\ell = 10$. In this implementation the
    computational graph of the algorithm was constructed in python
    and processed with TensorFlow version 1.3.0~\cite{abadi2016tensorflow}. For each size
    depicted, the computation becomes memory limited (128 GB RAM) after the highest
    depth shown (depths around $31$ to $33$ for $7 \times 7$ qubits depending
    on the instance).

\begin{figure}
  \centering
  \includegraphics[width=\columnwidth]{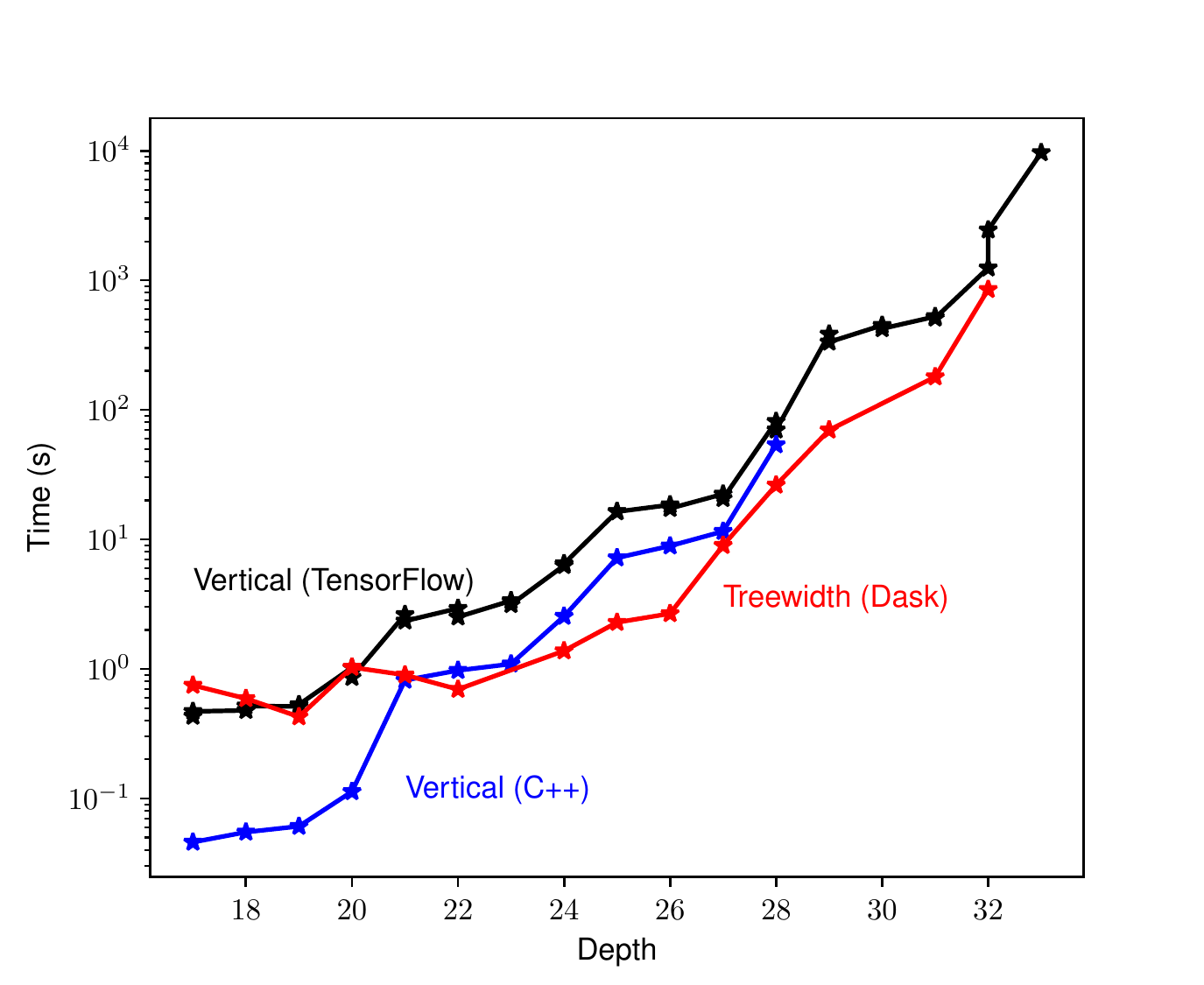}
  \caption{Timings per output probability for typical instances at
    different depths using different implementations of the graphical
    model variable elimination algorithm. We compare a TensorFlow
    implementation and a C++ implementation using the vertical variable
    elimination ordering (see Sec.~\ref{sec:sveo}), and an implementation using the Dask python
    package using a variable ordering obtained by running QuickBB~\cite{gogate2004complete} for a day to
  find a treewidth decomposition at each depth (see Sec.~\ref{sec:treewidth}).}
  \label{fig:timings_7_7}
\end{figure}

Figure~\ref{fig:timings_7_7} compares the run times per output
probability as a function of depth for three different
implementations. One is the implementation with a TensorFlow~\cite{abadi2016tensorflow} backend
using the same vertical variable elimination ordering as in
Fig.~\ref{fig:timings_sizes}. This implementation is efficient and
portable to the different architectures in which TensorFlow is
available, including distributed environments. A second implementation
is a native C++ implementation. The C++ implementation is faster,
although the run times are similar, and is probably less portable and harder to
distribute across machines. Finally we also show an implementation using a variable
ordering obtained by running QuickBB~\cite{gogate2004complete}  for a day to
  find a treewidth decomposition at each depth. The tensors
  obtained by this variable elimination ordering are worse aligned
  than in the vertical ordering, and can not be processed by the
  off-the-shelf TensorFlow binary. Therefore we used the python module
  Dask~\cite{Dask17} in this case (Dask was slower than TensorFlow in the vertical
  ordering for our current implementation). Not taking into account
  the time required to find a better variable elimination ordering,
  the treewidth ordering is faster. In all three cases further
  optimizations are possible, and the run times reported are merely
  indicative. All three implementations are memory (RAM) limited after
  the last depth shown in the figure.

\begin{figure}
  \centering
  \includegraphics[width=\columnwidth]{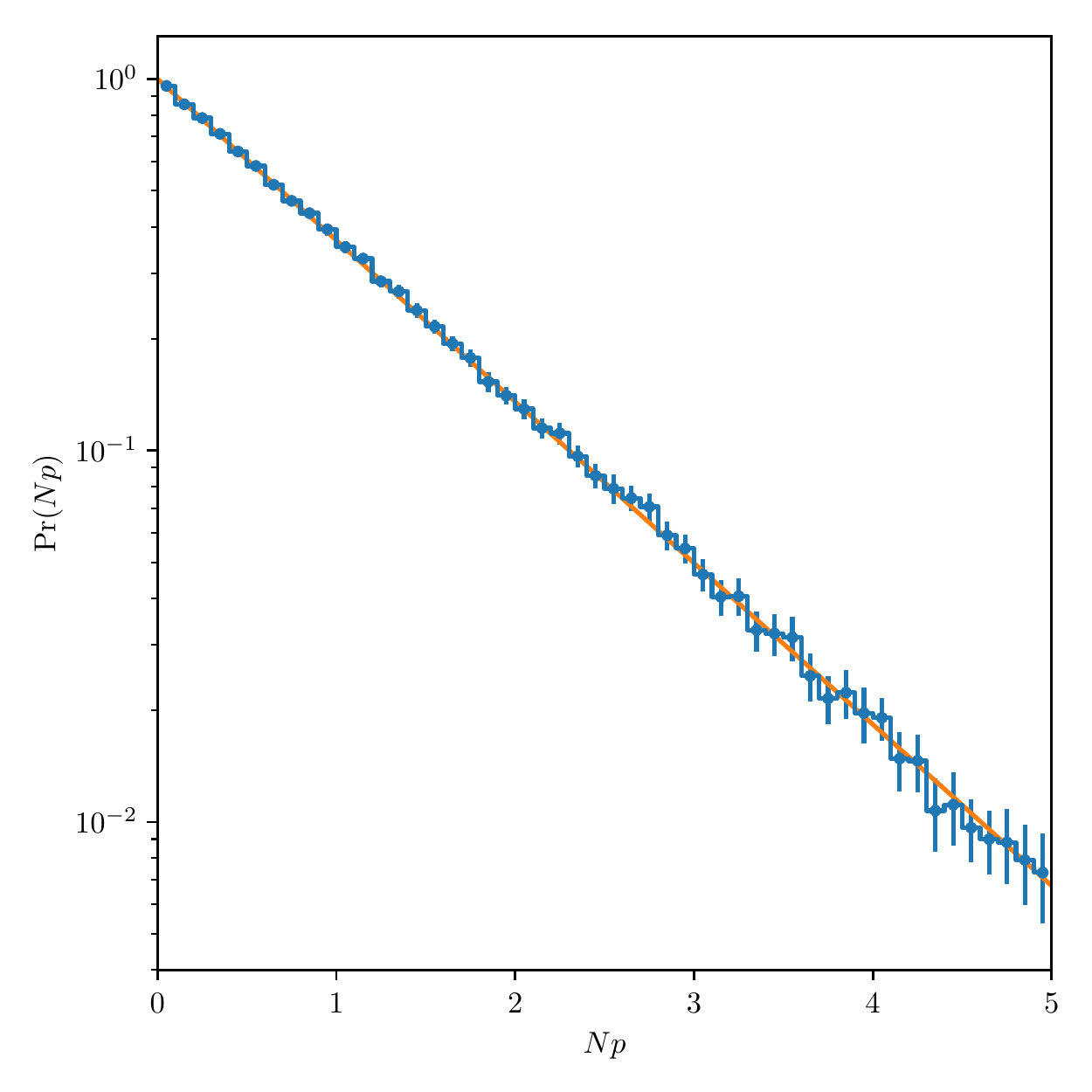}
  \caption{Histogram (blue) of the output probabilities for a circuit with $7
    \times 8$ qubits and depth 30. We calculated 200 thousand
    exact probabilities. Error bars were obtained by bootstrapping. We see that the probabilities obey the exponential or
    Porter-Thomas distribution (red)~\cite{boixo2016characterizing}.}
  \label{fig:pt}
\end{figure}

As an illustration, we calculated 200 thousand output probabilities of
a universal random circuit with $7 \times 8$ qubits and depth 30. We
show in Fig.~\ref{fig:pt} that the distribution obeys the exponential
or Porter-Thomas distribution. We also used these probabilities to
estimate the entropy of the output distribution of this circuit. We
obtain an entropy of $\log(2^{7\times8})-0.422 \pm 0.007$, while the analytical value
for a Porter-Thomas distribution is $\log(2^{7\times8}) - 0.4228$~\cite{boixo2016characterizing}.

\section{Circuit simulation as an undirected graphical model}\label{sec:ugm}

The basic principle of the algorithm can be seen as a mapping of a
quantum circuit to an undirected graphical model with complex
factors. This is equivalent to the interaction graph of the Ising
model detailed in Ref.~\cite{boixo2016characterizing}. The evaluation
is carried out using an algorithm developed in the context of exact
inference for undirected graphical
models~\cite{dechter1998bucket,murphy_machine_2012,bishop_pattern_2006}, but in the present
case the factors are complex. A
similar algorithm was used to find an exact ground state of a
classical Ising model in~\cite{boixo2014evidence}. We now explain in detail how the
undirected graphical model is constructed.

\subsection{Feynman path method}\label{sec:fp}

We represent a quantum circuit by a product of unitary matrices $U^{(t)}$ corresponding to different clock cycles $t$. We introduce the following notation for the amplitude of a particular bit-string after the final cycle of the circuit, 
\begin{equation}
 \langle x|\cU|0\ldots0\rangle=\sum_{\{b^{t}\}}\prod_{t=0}^{d-1}\bra{b^{t+1}}U^{(t)}\ket{b^{t}} ,\quad \ket{b^{d}}=\ket{x}.\label{eq:fpath}
\end{equation}
Here $\ket{ b^t}=\otimes_{j=1}^{n}\ket{b^t_j}$ and $\ket{b^t_j}$
corresponds to the states $\ket{ 0}$ or $\ket{1}$ of the  $j$-th
qubit.  The expression (\ref{eq:fpath}) can be viewed as a Feynman path integral with individual paths $\{b^{0},\ldots,b^{d}\}$
formed by a sequence of the computational basis states of the $n$-qubit system. The  initial condition for each path  corresponds to $b^{0}_j=0$ for all qubits and the final point corresponds to  
$\ket{b^{d}}=\ket{x}$. 

Assuming that a  $\T$ gate is applied to qubit $j$ at the cycle $t$, the indices of the matrix $\bra{b^{t+1}}U^{(t)}\ket{b^{t}}$ will be equal to each other, i.e. $b_j^{t+1}=b^{t}_j$. 
A similar property applies to the $\CZ$  gate as well. The state  of a qubit can only flip under the action of the gates $\H$, $\X^{1/2}$ or $\Y^{1/2}$.
We refer to these as non-diagonal gates as they contain two nonzero
elements in each row and column (unlike $\T$ and $\CZ$). This
observation allows us to rewrite the path integral representation in a
more economic fashion.

Through the circuit, a sequence of non-diagonal gates are applied to qubit $j$. We denote the length of this sequence as $d_j$.
In a given  path  the qubit $j$ goes through the  sequence of Boolean
states $\{b_j^k\}_{k=0}^{d_j},$  where,  as before, we have $b_j^k \in
\{0,1\}$.  The value of  $b_j^k$ in the sequence determines the state of the qubit  {\it immediately after}  the action of the $k$-th non-diagonal gate. The last element in the sequence is fixed   
by the assignment of bits in the bit-string $x$,
\begin{equation}
b_{j}^{d_j}=x_j\;,\quad j\in [1, n]\;.\label{eq:bc}
\end{equation}
Therefore, an individual path in the path integral can  be encoded by
the set of $\sum_{j=1}^{n}d_j$ binary variables $\{b_j^k\}$ with
$j\in[1, n]$ and  $k\in[0 , d_j]$ for a given $j$.

\subsection{Undirected graphical model}\label{subsec:ugm}

Equation~\eqref{eq:fpath} is a sum of products of factors defined by the gates of the quantum circuit.  A quantum circuit $\cU$ is given as a sequence of gates $\cU =U_g\cdots
U_1$. In expression~\eqref{eq:fpath}, each gate $U$ will contribute
factors defined by a complex pseudo-Boolean function
$\psi_{U}$. A diagonal one-qubit gate is expressed as
\begin{align}
  U = \sum_{b\in \{0,1\}} \psi_U(b) \ket {b} \bra {b}\quad\quad\textrm{(diagonal)}\;,
\end{align}
where $\psi_U(b)$ is a complex function of one Boolean variable. 
Next consider a non-diagonal one-qubit gate,
for example $\H$, $\X^{1/2}$, $\Y^{1/2}$, and so on,
\begin{align}
  U = \sum_{b,b' \in \{0,1\}} \psi_U
(b',b) \ket{b'} \bra{b}\quad\quad\textrm{(non-diagonal)}\;,
\end{align}
where $\psi_U(b,b')$ is a complex function of two Boolean
variables. For example, for a Hadamard gate $\psi_\H(1,1) = -1/\sqrt
2$, while all other entries are $1/\sqrt 2$. A
diagonal two-qubit gate, such as $\CZ$, can be written as
\begin{align}
  U=\CZ= \sum_{b,b' \in \{0,1\}}\psi_\CZ(b,b' )  \ket{b,b'}\bra{b,b'}\;,
\end{align}
where $\psi_\CZ(b_1,b_2 ) $ is also a function of two Boolean
variables, with $\psi_\CZ(1,1) = -1$ and all the other
entries equal $1$.

For circuits where the only two-qubit gate is a
$\CZ$, the expression~\eqref{eq:fpath} is a sum of products of factors
defined by complex functions of one or two Boolean variables
\begin{multline}
  \langle x|\cU|0\ldots0\rangle = \sum_{\{b_j^k \}}
  \delta(b_n^{d_n},x_n)\cdots
   \psi_\nu(b_{\nu_1}^{\nu'_1} ,b_{\nu_2}^{\nu'_2})
\cdots  \\ 
   \cdots\psi_\mu(b_\mu^{\mu'}) \cdots\psi_1(b_{1}^1,b_1^0)\cdots \delta(b_2^0,0)\delta(b_1^0,0)
  \label{eq:psi_x}
\end{multline}
where $\{b_j^k\}$ are the classical Boolean variables introduced
above. The delta functions set the values of the Boolean variables
corresponding to the initial and final bit-strings. If we omit some
or all the
delta functions, we would compute sets of amplitudes at once. We show below how the method is extended to more general
gates or gates acting on more qubits. 

In order to compute expression $\eqref{eq:psi_x}$, it is convenient
to interpret it as an undirected
graphical model by defining the graph $G$ where each variable $b_j^k$
corresponds exactly to one vertex in $G$ and each function $\psi$ results in a
clique between the vertices corresponding to the variables of
$\psi$. As explained above, the subscript $j \in [1,n]$ enumerates the
qubits, and the superscript $k$ enumerates new variables introduced
along each qubit worldline, one new variable for each qubit acted upon
by a non-diagonal gate. After the vertices of the graph $G$ have been
defined in this way, gates do not introduce any further vertices,
rather they (might) introduce edges between vertices. Multiple
gates might be represented by the same edge or edges, and act upon the
same vertex or vertices. Explicit rules on how gates are represented
are given below. 

A quantum circuit representation of a diagonal one-qubit gate and the corresponding graphical model are
  \begin{align}
    \Qcircuit@C=1em@R=.7em@!R{
      &\ustick{b} \qw & \gate{U} & \ustick{b} \qw &\qw
  } \quad\quad \Longrightarrow \quad\quad
  \begin{tikzpicture}
    \node (b) at (0,0) [circle,draw, label=above:$b$] {};
  \end{tikzpicture}\;.
  \end{align}
The quantum circuit and graphical model of a non-diagonal one-qubit gate
are
  \begin{align}
    \Qcircuit@C=1em@R=.7em@!R{
      &\ustick{b} \qw & \gate{U} & \ustick{b'} \qw &\qw
  } \quad\quad \Longrightarrow \quad\quad
  \begin{tikzpicture}
    \node (bi) at (0,0) [circle,draw, label=above:$b$] {};
    \node (bo) at (1,0) [circle,draw, label=above:$b'$] {}; 
    \draw (bi) -- (bo) {}; 
  \end{tikzpicture}\;.
  \end{align}
The circuit and graphical model for a two-qubit diagonal gate are
  \begin{align}
    \begin{aligned}
      \Qcircuit@C=1em@R=2em@!R{
        &\ustick{b} \qw & \multigate{1}{U} & \ustick{b} \qw &\qw \\
        &\ustick{b'} \qw & \ghost{U} & \ustick{b'} \qw &\qw \\
      }
    \end{aligned}
\quad\quad \Longrightarrow \quad\quad
\begin{aligned}
  \begin{tikzpicture}
    \node (b1) at (0,.5) [circle,draw, label=west:$b$] {}; 
    \node (b2) at (0,-.5) [circle,draw, label=west:$b'$] {}; 
    \draw (b1) -- (b2) {};
  \end{tikzpicture}\;.
\end{aligned}
\end{align}
Finally, the corresponding quantum circle and graphical model for a two-qubit non-diagonal gate  are
 \vspace{.2cm} \begin{align*}
    \begin{aligned}
      \Qcircuit@C=1em@R=2em@!R{
        &\ustick{b_0} \qw & \multigate{1}{U}& \ustick{\;\;b_0'} \qw &\qw \\
        &\ustick{b_1} \qw & \ghost{U} & \ustick{\;\;b_1'} \qw &\qw \\
      }
    \end{aligned}
\quad\quad \Longrightarrow \quad\quad
\begin{aligned}
  \begin{tikzpicture}
    \node (bi1) at (0,.5) [circle,draw, label=west:$b_0$] {}; 
    \node (bi2) at (0,-.5) [circle,draw, label=west:$b_1$] {}; 
    \node (bo1) at (1,.5) [circle,draw, label=east:$b_0'$] {}; 
    \node (bo2) at (1,-.5) [circle,draw, label=east:$b_1'$] {}; 
    \draw (bi1) -- (bi2) -- (bo2) -- (bo1) -- (bi1) -- (bo2) {};
    \draw (bi2) -- (bo1) {};
  \end{tikzpicture}\;.
\end{aligned}
  \end{align*}

As an example, consider a quantum circuit with two qubits such as
  \begin{align}
    \begin{aligned}
      \Qcircuit@C=1em@R=1em@!R{
        \lstick{\ket{0}}& \gate{H} & \ctrl{1} & \gate{H} & \meter \\
        \lstick{\ket{0}}& \gate{H} & \ctrl{-1} &  \gate{H}  &\meter \\
      }
    \end{aligned}\label{eq:ex_circ}
  \end{align}
where the vertical line symbolizes a controlled-Z gate (which is
diagonal) and the final boxes represent a measurement. 
The undirected graphical model corresponding to this circuit is
\begin{align}
\begin{aligned}
  \begin{tikzpicture}
    \node (a0) at (0,1) [circle,draw] {};
    \node (b0) at (1,1) [circle,draw] {};
    \node (c0) at (2,1) [circle,draw] {}; 
    \node (a1) at (0,0) [circle,draw] {};
    \node (b1) at (1,0) [circle,draw] {};
    \node (c1) at (2,0) [circle,draw] {};
    \draw (a0) to (b0) to (c0); 
    \draw (a1) to (b1) to (c1); 
    \draw (b1) to (b0);
  \end{tikzpicture}\;.
\end{aligned}\label{fig:ex_ugm}
  \end{align}
This example is worked out explicitly in App.~\ref{app:example}.

The same method can be applied to calculate the expectation value of
an operator $O$ given by $\bra 0 \cU^\dagger O \cU \ket 0$. For a local
operator we simplify this expression by writing the circuit unitary
$\cU$ in terms of gates $\cU_\alpha$, and canceling terms
$\cU_\alpha^\dagger \cU_\alpha = 1$ whenever possible. The operator
results in factors in an expression equivalent to
Eq.~\eqref{eq:psi_x}, following exactly the same
procedure. Alternatively, the expectation value of an observable can
be estimated by sampling output probabilities and using Monte Carlo
integration, see App.~\ref{app:mcerror}.

\section{Variable elimination algorithm}\label{sec:ve}

The evaluation of a sum of products of factors such as expression
$\eqref{eq:psi_x}$ can be done directly by  an algorithm developed in
the context of exact inference for undirected graphical models, known as the bucket elimination algorithm~\cite{dechter1998bucket,kask2001bucket} or the variable elimination algorithm~\cite{murphy_machine_2012}. In our
case, though, the factors have complex values, and it does not
correspond to a probabilistic model. The
graphical model of some circuits can also be interpreted directly as
an Ising model at imaginary
temperature~\cite{boixo2016characterizing}, see App.~\ref{app:pf},
similar to the relation between undirected graphical models or random Markov fields and Ising models at real temperature. 

The variable elimination algorithm is usually explained by first considering the case where the graphical model is one dimensional. This would correspond to the worldline of a quantum circuit with a single qubit. In this case we would have to evaluate expressions of the form
\begin{align}
  &\sum_{b^{d_0}, \cdots ,b^0} \psi_{d_0}(b^{{d_0}}, b^{{d_0}-1}) \cdots\psi_2(b^2,b^1) \psi_1(b^1,b^0 ) \nonumber \\
  &=\sum_{b^{d_0},b^{{d_0}-1}} \psi(b^{{d_0}}, b^{{d_0}-1}) \cdots \sum_{b^{1}} \psi_2(b^2,b^1) \sum_{b^0} \psi_1(b^1,b^0 ) \nonumber \\
  &=\sum_{b^{d_0},b^{{d_0}-1}} \psi(b^{{d_0}}, b^{{d_0}-1}) \cdots \sum_{b^{1}} \psi_2(b^2,b^1) \tau_1(b^{1})\label{eq:1delim1} \\
  &=\sum_{b^{d_0},b^{{d_0}-1}} \psi(b^{{d_0}}, b^{{d_0}-1}) \cdots \sum_{b^{2}} \psi_2(b^3,b^2)\tau_2(b^{2})\label{eq:1delim2}  \\
  &=\sum_{b^{d_0}} \tau_{d_0}(b^{d_0})\;. \nonumber
\end{align}
That is, we evaluate a one dimensional undirected graphical model by
using the distributive property and eliminating variables from left to
right (or right to left). A variable is eliminated by performing the
corresponding sum and storing the corresponding factor in memory, as
in the step from Eq.~\eqref{eq:1delim1} to Eq.~\eqref{eq:1delim2}. The
same algorithm applies to trees, and is known as belief propagation in
the context of graphical models~\cite{pearl1982reverend}. It is also
closely related to the  Bethe Peierls iterative
method~\cite{bethe1935statistical} or the cavity method in physics~\cite{mezard1987spin}. The cost of this algorithm is linear.

In more general cases we obtain factors that grow in size after eliminating a variable. For example
\begin{align}
  \sum_{b_j} \psi(b_i, b_j) \psi(b_j,b_k) \tau(b_j,b_l) = \tau(b_i,b_k,b_l)\;.\label{eq:par_sum}
\end{align}
The size of a factor or tensor stored in memory is exponential in the
number of variables or indexes. The size of the factors obtained through variable elimination depends on the order of elimination. Consequently, the cost of this algorithm can vary exponentially for different variable elimination orders.

\subsection{Vertical variable elimination ordering for low-depth
  circuits}\label{sec:sveo}
For circuits with low depth and low dimension a simple strategy for
the order of variable elimination is to process one qubit at a time,
eliminating all variables in one
worldline sequentially before moving to a neighboring qubit.  As an example, we consider again the
circuits of Ref.~\cite{boixo2016characterizing}, which are defined in
a two dimensional lattice of qubits. As explained in
Sec.~\ref{sec:ugm}, the mapping of a circuit output amplitude to an undirected
graphical model results in a graph defined on vertices corresponding
to Boolean classical
variables $b_j^k$, where the index $j$
enumerates the qubits, and the superscript $k$ enumerates new
variables along the so-called worldline of a qubit $j$ in the time
direction. We assume that the qubit index $j$ is ordered so that
sequential values correspond to neighboring qubits in the underlying
two dimensional lattice. Processing the variables first along the
worldline direction, which we call the vertical ordering of variable
elimination, corresponds to eliminating variables in the
lexicographical order of the pairs $(j,k)$. That is, we eliminate all
$b_j^k$ variables corresponding to qubit $j$ sequentially along the $k$
index before moving to the variables corresponding to qubit $j+1$. 

\begin{figure}
  \centering
  \includegraphics[width=\columnwidth]{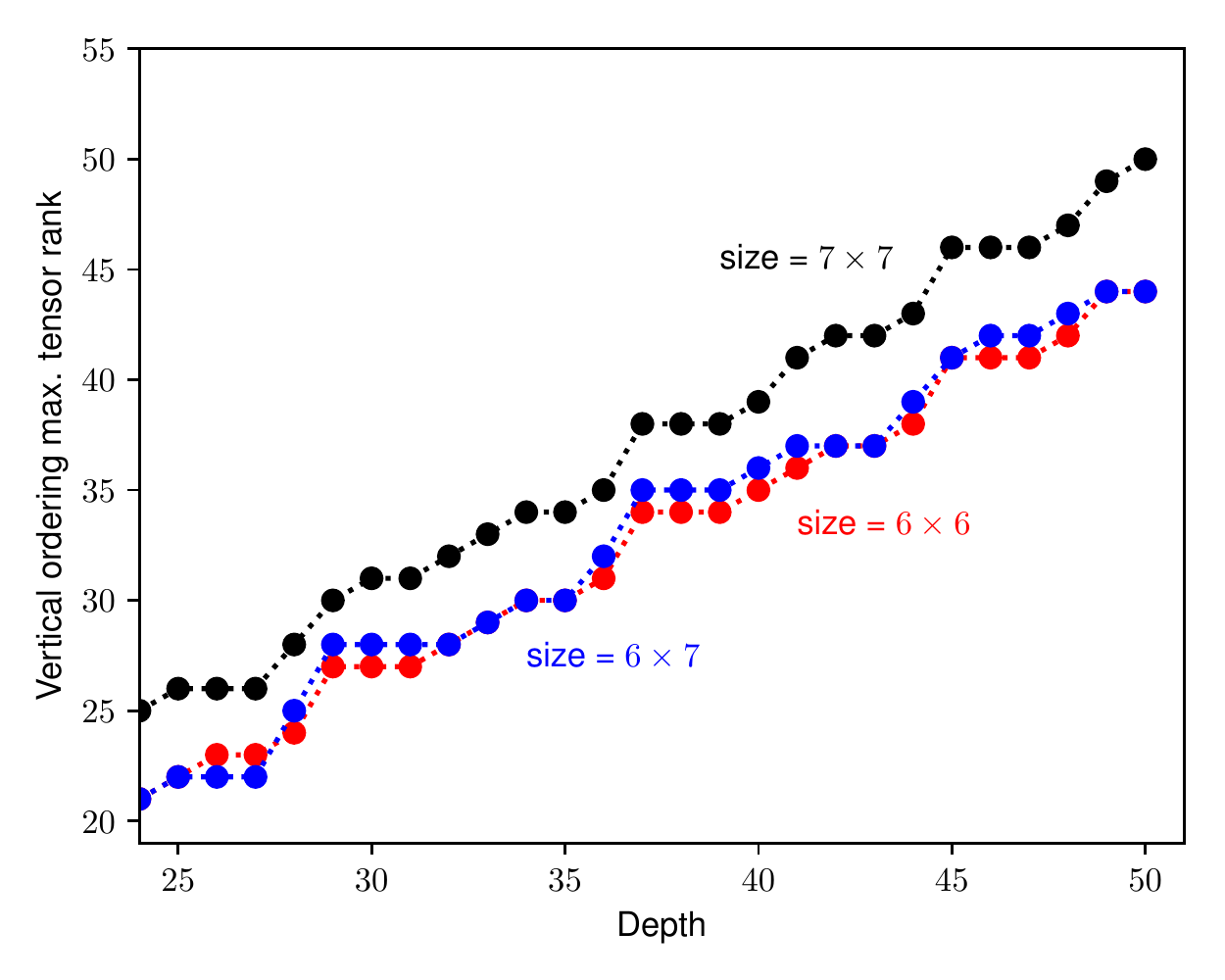}
  \caption{Treewidth for circuits from
    Ref.~\cite{boixo2016characterizing} with $6 \times 6$, $7 \times
    6$, and $7 \times 7$ qubits as a function of the circuit depth
    using a vertical elimination ordering, one circuit instance in each case. There are small variations
    between instances of the same size.}
  \label{fig:treewidth_vertical}
\end{figure}

Figure~\ref{fig:treewidth_vertical} shows the size of the maximun
tensor rank as a function of the circuit depth for an instance of each
size $6 \times 6$, $6 \times 7$, and $7 \times 7$ using a vertical
elimination ordering. More exactly, we plot the size of the tensor
minus 1, to make it directly comparable to the standard definition of
treewidth used in Fig.~\ref{fig:treewidth}.
There are small variations between instances of
the same size, which explains that in this particular case the
instance of size $6 \times 7$ has a larger tensor size for some depths in this
ordering than the instance of size $7 \times 7$.

\subsection{Numerical estimation of the treewidth for some quantum
  circuits}\label{sec:treewidth}

The size of the factors obtained by variable elimination from a given ordering can be analyzed graphically. We start with the undirected graph corresponding to the original expression, where each variable corresponds to a vertex. To simulate the elimination of a variable, we add an edge between all vertices connected to the vertex (variable) being eliminated. The cliques obtained in this process correspond to factors obtained through variable elimination. The size of the largest clique is called the induced width. Note that different orderings in the vertex elimination process, which correspond to different orderings of variable elimination, can result in different  cliques.  The treewidth is defined as the minimum width over all variable orderings. Determining the treewidth is in general NP-complete, but heuristic algorithms, such as QuickBB~\cite{gogate2004complete}, can be used to obtain an approximation. 

\begin{figure}
  \centering
  \includegraphics[width=\columnwidth]{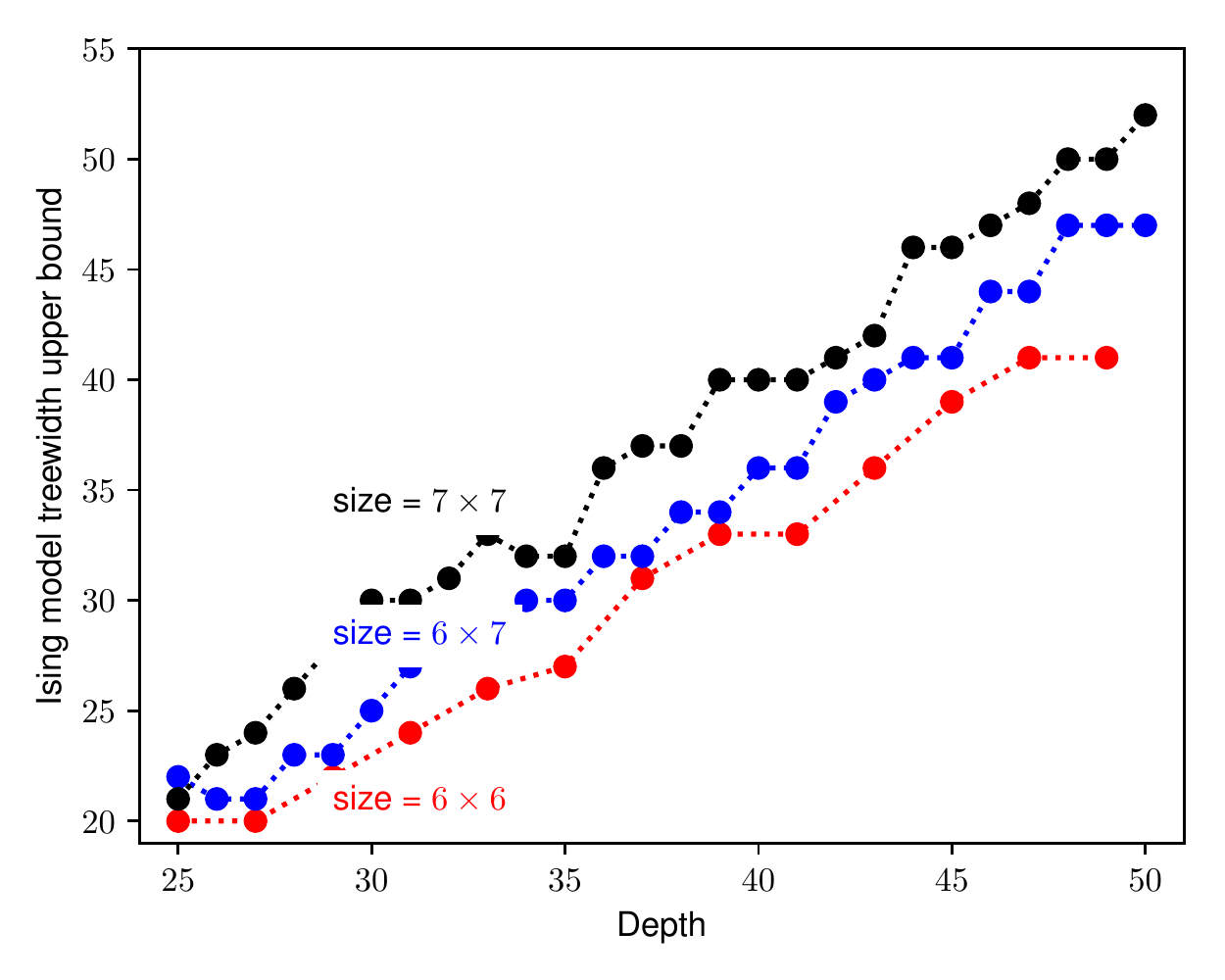}
  \caption{Numerical upper bound for the treewidth of the undirected graphical model corresponding to circuits from Ref.~\cite{boixo2016characterizing} with $6 \times 6$, $6 \times 7$, and $7 \times 7$ qubits as a function of the circuit depth.}
  \label{fig:treewidth}
\end{figure}

For a circuit in a 2D lattice of qubits with two-qubit gates
restricted to nearest neighbors, as in
Ref.~\cite{boixo2016characterizing}, the treewidth of the
corresponding undirected graphical model is proportional to
$\min(O(d \ell), O(n))$, where $\ell$ is the minimum lateral
dimension. Figure~\ref{fig:treewidth} shows numerical upper bounds for
the treewidth as a function of depth (see also
Ref.~\cite{boixo2016characterizing}). The upper bounds were obtained
by running the \emph{QuickBB}
algorithm~\cite{gogate2004complete}. This algorithm also returns the
corresponding variable elimination order, which was used in
Fig.~\ref{fig:timings_7_7} and also to calculate the probabilities plotted
in Fig.~\ref{fig:pt}.

 \section{Conclusion}
 We generalize the simulation of a quantum circuit by mapping it onto
 an undirected graphical model, which we compute using the variable
 elimination algorithm, developed in the context of exact
 inference~\cite{dechter1998bucket,bishop_pattern_2006,murphy_machine_2012}.  This is particularly convenient for circuits of
 low depth~\cite{markov_simulating_2008,boixo2016characterizing,aaronson2016complexity}, and with predominantly diagonal two-qubit
 gates. We are able to sample the output distribution of quantum
 circuits in a single workstation for cases that previously required a
 supercomputer.  We also show how to perform cross entropy benchmarking and
 how to estimate the value of an
 observable. The computation of many amplitudes is embarrassingly
 parallelizable across machines. The cost of this algorithm is
 exponential in $\min(O(d \ell), O(n))$ for depth $d$, minimun lateral
 dimension $\ell$ and total number of qubits $n$. More optimally, the
 cost is exponential in the treewidth of the graph, as estimated in
 Ref.~\cite{boixo2016characterizing}. Follow up work will study
 potential improvements to this algorithm, such as implementation in a
 supercomputer and code optimization, and possible space-time
 tradeoffs~\cite{aaronson2016complexity}.

\begin{acknowledgments}
  \label{sec: Acknowledgment}
  We acknowledge Stephan Hoyer for conversations and suggesting the use of Dask. 
\end{acknowledgments}

\appendix

\section{Pseudocode}\label{app:pseudocode}

In this section we give pseudocode for the quantum circuit simulation
algorithm presented in the main text~\cite{murphy_machine_2012}
. As seen
in Eq.~\eqref{eq:psi_x}, an output amplitude of a quantum circuit is expressed as a
sum of products of factors. Each factor is a complex valued function
or tensor defined in one or two Boolean
variables or indexes. The variable elimination algorithm performs the sum of
Eq.~\eqref{eq:psi_x} eliminating one variable at a time, see
Eq.~\eqref{eq:par_sum}. Larger tensors are created in the process. In
what follows we assume that all Boolean variables have been given an
appropriate order for variable elimination, see
Secs.~\ref{sec:sveo} and~\ref{sec:treewidth}.

The basic objects of the computation are
\lstinline{tensors}. \lstinline{Tensors} are complex functions defined
on Boolean variables. Variables are associated with
vertices in the undirected graphical model as explained in
Sec.~\ref{subsec:ugm}. \lstinline{Tensors} implement two
operations. One is a multiplication across tensors, \lstinline{tensor1 * tensor2}, implemented with broadcasting semantics, as in the NumPy
python library. This means that the product of \lstinline{tensors} is done
elementwise. If an index or variable in \lstinline{tensor_a} is
missing in \lstinline{tensor_b} then, in effect, this index is added
to \lstinline{tensor_b} with dimension 1
(the value of \lstinline{tensor_b} is determined by the pre-existing indexes). The other operation,
\lstinline{tensor.reduce_sum()}, computes the sum of elements across the
first dimension of the \lstinline{tensor}. For simplicity, we assume
that the indexes of all \lstinline{tensors} are initially ordered from low to
high. This will also be an invariant of the algorithm. 

The bucket elimination
algorithm~\cite{dechter1998bucket,kask2001bucket} keeps track of the
order of the operations involved. More explicitly, we use the bucket
elimination algorithm to build a computational graph that will be
processed by TensorFlow or Dask. A \lstinline{bucket} is
a list of \lstinline{tensors}. The ordered list \lstinline{buckets}
has one \lstinline{bucket} per variable. In this way each \lstinline{bucket} is associated to a
variable. An invariant of the algorithm is that each \lstinline{tensor} in a
\lstinline{bucket} acts upon the variable associated with the
\lstinline{bucket}, and all other variables have higher order.

First we initialize each \lstinline{bucket} with the
\lstinline{tensors} in Eq.~\eqref{eq:psi_x}, corresponding to the
gates, initialization and measurement of a circuit output amplitude:
\begin{lstlisting}
for tensor in initial_factors:
  buckets[tensor.indexes[0]].append(tensor)
\end{lstlisting}
Note that this obeys the invariant given above, because of the
assumption that the variables or indexes in
each \lstinline{tensor} are ordered from low to high.

Each \lstinline{bucket} in the \lstinline{buckets} list is processed
in order. Processing a bucket implements the elimination of the
variable associated with that bucket, as in Eq.~\eqref{eq:par_sum}. As
noted above, we assume that variables have initially been ordered with
the chosen variable elimination order. Processing the
\lstinline{buckets} list one by one then implements the correct
order. Processing a \lstinline{bucket} can return a \lstinline{tensor}
without indexes, corresponding to a scalar, when all tensors in the
\lstinline{bucket} had only one variable (this has to be the variable
associated with the bucket). Typically this only happens in the last
\lstinline{bucket}, and this scalar corresponds to the amplitude that
we want to calculate. In some cases, for circuits of very small depth,
the circuit can be decomposed into disjoint tensors, and the
corresponding scalars are multiplied together.
\begin{lstlisting}
result = None
for bucket in buckets:
  if not empty(bucket):
    tensor = bucket.process_bucket()
    if rank(tensor) > 0:
      first_index = tensor.indexes[0]
      to_bucket = buckets[first_index]
      to_bucket.append(tensor)
    else:
      if result:
        result *= tensor
      else:
        result = tensor
return result
\end{lstlisting}

The processing of a \lstinline{bucket} consists on
two steps: first all \lstinline{tensors} in the \lstinline{bucket} are broadcasted together, and then the
first variable is summed over, as in Eq.~\eqref{eq:par_sum}. By the invariants of the algorithm, the
variable summed over is the one associated to the
\lstinline{bucket}.
\begin{lstlisting}
def process_bucket(bucket):
  result = bucket.tensors[0]
  for tensor in bucket.tensors[1:]:
    result *= tensor
  # reduce
  return result.reduce_sum()
\end{lstlisting}

\section{Cross entropy benchmarking}\label{app:ceb}
We now review cross entropy benchmarking, a technique to  estimate the fidelity of
an experimental quantum circuit
implementation using the exact probabilities amplitudes calculated by
a quantum circuit simulation algorithm~\cite{boixo2016characterizing, neill2017}. 

The cross-entropy between a probability distribution $p_A$ and the ideal output probability $p_\cU$ is defined as
$S(p_A, p_\cU) = -\sum_x p_A(x) \log p_\cU(x)$. As explained
in Ref.~\cite{boixo2016characterizing}, the cross-entropy with an
uncorrelated distribution $p_{\rm unc}$ averaged over quantum
circuits is almost constant, $\H_0 = \mE_\cU[ S(p_{\rm unc} , p_U)]$. In the
following we make the averaging over circuits $\cU$ implicitly. 

We write the output of an experimental quantum circuit
implementation with fidelity $\alpha$ as
  \begin{align}
  \rho = \alpha\,\cU \ket {0} \!\bra {0} \cU^\dagger + (1-\alpha) \sigma_\cU\;.
\end{align}
 The state $\cU
 \ket {0}$ corresponds to the ideal output of circuit $\cU$, without
 any errors, while $\sigma_\cU$
 represents the output of the implementation of $\cU$ after any
 experimental errors. The experimental output probability for a bit-string $x$ is
 $p_\exp(x) = \bra{x} \rho\ket{x} = \alpha \, p_\cU(x) +
 (1-\alpha) \bra x \sigma_\cU \ket x$. From the arguments
 presented in Ref.~\cite{boixo2016characterizing}, almost  any error in a
 random universal circuit of sufficient depth results in an output probability
 which is uncorrelated with the ideal probability distribution, and
 therefore we assume that $\bra x \sigma_\cU
 \ket x$ is uncorrelated with $p_\cU(x)$. 
We have
\begin{align}
      &\H_0 - S(p_\exp, p_\cU)  \\
             & = \H_0 + \sum_j\big( \alpha  p_\cU(x_j) \nonumber
               \\ &\quad\quad+  (1-\alpha) \bra {x_j}  \sigma_\cU \ket {x_j}\big) \log p_\cU(x_j) \\
      & = \H_0 - \alpha \H(p_\cU) - (1-\alpha) \H_0 \\ &= \alpha(\H_0 - \H(p_\cU))\;.
    \end{align}
Therefore
\begin{align}\label{eq:ae}
    \alpha = {\H_0 - S(p_\exp, p_\cU) \over \H_0 - \H(p_\cU)}
\end{align}
is an estimate of the experimental fidelity. Note that $\H_0$ and
$\H(p_\cU)$ can be calculated numerically, see Sec.~\ref{sec:nums} and
App.~\ref{app:mcerror}. Furthermore, for universal random circuits
in the Porter-Thomas regime we have $\H_0 = \log(2^n) + \gamma$ and
$\H(p_\cU) = \log(2^n) -1 + \gamma$, where $\gamma$ is Euler's
constant~\cite{boixo2016characterizing}. 

In order to estimate the fidelity $\alpha$ of an experimental
implementation, we take a large sample
$S_\exp=\{x_{1}^\exp,\ldots,x_{m}^\exp\}$ of bit-strings $x$ in the
computational basis ($m \sim 10^3 - 10^6$). From the central limit
theory we have (see App.~\ref{app:mcerror})
\begin{align}
  &- \frac 1 m \sum_{j =1}^{m} \log          p_U(x_j^\exp ) \label{eq:cee}\\ &= -\sum_x
                                                                 p_\exp(x) \log p_\cU(x) + O(1/\sqrt m) \\
  &= S(p_\exp, p_\cU)+ O(1/\sqrt m)\;.
\end{align}
In conclusion, using the algorithm presented in the main text to
calculate the $m$ ideal output probabilities $\{p_\cU(x^\exp)\}$, we can use
Eqs.~\eqref{eq:cee} and \eqref{eq:ae} to obtain an estimate of
$\alpha$. 

\section{Random circuit sampling with a set of
  probabilities}\label{app:mcerror}
Assume that we are asked to perform the following computational task:
sample a set of bit-strings $S$ from the output distribution
$\{p_\cU(x)\}$ defined by a quantum circuit $\cU$.
In order to simulate sampling from the output distribution $\{p_\cU(x)\}$ using the
algorithm above, we start with a set $T = \{x_j\}$ of bit-strings 
selected uniformly at random. We denote the size of $T$ as $t$. We then calculate the probabilities $\{p_\cU(x_j)\}$ corresponding
to the circuit $\cU$ and sample from the set $T$ according to the normalized probabilities $\tilde{p}_\cU(x_j)$
\begin{align}
\tilde{p}_\cU(x_j) &= {p_\cU(x_j) \over \sum_{j \in T} p_\cU(x_j)} = \frac
{2^n} t\(1+O\( {1\over \sqrt t}\) \)p_\cU(x_j) \;.
\end{align}
Note that by standard arguments in Monte Carlo integration, and given
that the variance of the relevant distribution for random circuits is a constant $\sim
1$~\cite{boixo2016characterizing}, we can approximate any function
defined over the output probabilities of the circuit using the set
$T$ up to an error $\sim 1/\sqrt{t}$. 

In particular, if we take a sample $S$ of size $m$ from $T$ according
to the probabilities $\{\tilde p_\cU(x_j)\}$ we would get an estimate of
the cross entropy of this sample
\begin{align}
    &-\frac 1 m  \sum_{j \in S} \log p_\cU(x_j) \nonumber \\ &= -\sum_{j \in T} \tilde
  p_\cU(x_j) \log p_\cU(x_j) +O\(\frac 1 {\sqrt m}\) \nonumber \\  &= -\frac {2^n}
   t \(1+O\( {1\over \sqrt t}\)\)\sum_{j \in T} p_\cU(x_j) \log
                                                                   p_\cU(x_j) +O\(\frac 1 {\sqrt m}\) \nonumber  \\
              &= H(\cU) +O\( {1\over \sqrt t}\)H(\cU) +O\(\frac 1 {\sqrt m}\)\;,
\end{align}
where $H(\cU)$ is the entropy of the output distribution defined by
circuit $\cU$. 
That is, according to the cross entropy difference (see
Eq.~\eqref{eq:ae} and Ref.~\cite{boixo2016characterizing}), our simulation is
sampling from the ideal circuit with high fidelity, because we obtain
the correct entropy.

A more detailed calculation shows that the cross entropy obtained this
way is distributed according to 
\begin{multline}
  H(\cU)  - \xi \sqrt{2/t}\, H(\cU) + \zeta \sqrt{(\pi^2/6 -1)/m} \\+ \xi
  \zeta \frac{n^2}{\sqrt{2 t m (\pi^2/6 -1)}}\;,
\end{multline}
where $\xi$ and $\zeta$ and independent random variables that obey
the normal distribution $N(0,1)$.

 Note that the error or order $1/\sqrt{m}$ comes
from simulating a sample of size $S$ out of the set $T$ of output
probabilities that have been calculated. This error is avoided if we
estimate an observable directly from $T$.

\section{Simulation example}\label{app:example}
In this appendix we consider the specific example of the quantum
circuit
\begin{align}
    \begin{aligned}
      \Qcircuit@C=1em@R=1em@!R{
        \lstick{\ket{0}}& \gate{H}& \ctrl{1} & \gate{H} & \meter \\
        \lstick{\ket{0}}& \gate{H} & \ctrl{-1} &  \gate{H} &\meter \\
      }
    \end{aligned}\label{fig:app_circ}
  \end{align}
We calculate the output amplitude for input $\ket{00}$ and output
$\ket{00}$. The corresponding graphical model is
\begin{align}
\begin{aligned}
  \begin{tikzpicture}
    \node (a0) at (0,1.5) [circle,draw] {$b_0^0$};
    \node (b0) at (1.5,1.5) [circle,draw] {$b_0^1$};
    \node (c0) at (3,1.5) [circle,draw] {$b_0^2$}; 
    \node (a1) at (0,0) [circle,draw] {$b_1^0$};
    \node (b1) at (1.5,0) [circle,draw] {$b_1^1$};
    \node (c1) at (3,0) [circle,draw] {$b_1^2$};
    \draw (a0) to (b0) to (c0); 
    \draw (a1) to (b1) to (c1); 
    \draw (b1) to (b0);
  \end{tikzpicture}\;.
\end{aligned}\label{fig:ex_with_variables}
  \end{align}
Because the input and output is specified, we have $b_0^0$ $=$ $b_0^2$
$=$ $b_1^0$ $=$ $b_1^2$ $=$ 0. Then the graph can be simplified to 
\begin{align}
\begin{aligned}
  \begin{tikzpicture}
    \node (b0) at (1.5,1.5) [circle,draw] {$b_0^1$};
    \node (b1) at (1.5,0) [circle,draw] {$b_1^1$};
    \draw (b1) to (b0);
  \end{tikzpicture}\;.
\end{aligned}\label{fig:ex_shorted}
  \end{align}
The treewidth of this graph is 2, because it is a clique with two vertices.\footnote{This is the same as the
  treewidth of the graph in Fig.~\eqref{fig:ex_with_variables} when first
  eliminating the variables $b_0^0$, $b_0^2$, $b_1^0$ and
  $b_1^2$.} In the next step we can eliminate the variables $b_0^1$
and $b_1^1$ in any order. 

More explicitly, the amplitude $\bra{00} C \ket{00}$, where $C$ is the
circuit in Fig.~\eqref{fig:app_circ}, is
\begin{multline}
   \bra{00} C \ket{00} = \sum_{b_0^1,b_1^1}
   \psi_\H(0,b_0^1)\psi_\CZ(b_0^1,b_1^1) \psi_\H(b_0^1,0) \\
   \psi_\H(0,b_1^1) \psi_\H(b_1^1,0)\;.\label{eq:a1}
\end{multline}
The function $\psi_\H$ corresponds to a Hadamard gate and is given by
the table
\begin{center}
\begin{tabular}{ |cc|c| } 
 \hline
 0& 0& $1/\sqrt{2}$ \\ 
 0& 1& $1/\sqrt{2}$ \\ 
 1& 0& $1/\sqrt{2}$ \\  
 1& 1& $-1/\sqrt{2}$ \\ 
 \hline
\end{tabular}
\end{center}
The function $\psi_\CZ$ corresponds to a controlled-Z gate and is given by
the table
\begin{center}
\begin{tabular}{ |cc|c| } 
 \hline
 0& 0& $1$ \\ 
 0& 1& $1$ \\ 
 1& 0& $1$ \\  
 1& 1& $-1$ \\ 
 \hline
\end{tabular}
\end{center}

We can rewrite Eq.~\eqref{eq:a1} as
\begin{align}
   \bra{00} C \ket{00} = \sum_{b_0^1,b_1^1}
   \tau_1(b_0^1,b_1^1) \label{eq:a2}
\end{align}
where the table for the function $\tau_1$ is
\begin{center}
\begin{tabular}{ |cc|c| } 
 \hline
 0& 0& $1/4$ \\ 
 0& 1& $1/4$ \\ 
 1& 0& $1/4$ \\  
 1& 1& $-1/4$ \\ 
 \hline
\end{tabular}
\end{center}

If we now sum over the variable $b_1^1$ we obtain
\begin{align}
   \bra{00} C \ket{00} = \sum_{b_0^1}
   \tau_2(b_0^1)
\end{align}
where the function $\tau_2$ is 
\begin{center}
\begin{tabular}{ |c|c| } 
 \hline
 0& $1/2$ \\ 
 1& 0 \\ 
 \hline
\end{tabular}
\end{center}

Finally, summing $b_0^1$ we obtain
\begin{align}
    \bra{00} C \ket{00} = 1/2\;.
\end{align}

\section{Relation to tensor network contraction}\label{sec:tn}

Variable elimination on the undirected graph representing a quantum circuit is related to the simulation of a quantum circuit by contracting tensor networks~\cite{markov_simulating_2008}. 
In a tensor network graphical representation, gates correspond to vertices (tensors) and qubits connecting gates (indexes in the tensors) are represented by edges. Therefore, a two-qubit gate is represented as a node with four edges, two for input and two for output. For instance, the tensor network corresponding to the example of Eq.~\eqref{eq:ex_circ} is   \begin{align}
    \begin{tikzpicture}
      \node (a) at (0,1) [circle,draw] {}; 
      \node (a2) at (1,1) [circle,draw] {};
      \node (e) at (0,0) [circle,draw] {};
      \node (f) at (1,0) [circle,draw] {};
      \node (b) at (2,.5) [circle,draw] {};
      \node (c) at (3,1) [circle,draw] {};
      \node (c2) at (3,0) [circle,draw] {};
      \node (d) at (4,1) [circle,draw] {};
      \node (g) at (4,0) [circle,draw] {};
      \draw (a) to (a2);
      \draw (b) to [out=240,in=0] (f);
      \draw (b) to [out=125,in=0] (a2);
      \draw (e) to (f);
      \draw (b) to [out=50,in=180] (c);
      \draw (b) to [out=315,in=180] (c2);
      \draw (c) to (d);
      \draw (c2) to (g);
    \end{tikzpicture}\label{fig:ex_tn}\;.
  \end{align}

The operation of contracting a network of tensors is analogous to the variable elimination in Sec.~\ref{sec:ve}. The indexes in common are summed over or contracted, and the size of the resulting tensor in memory is exponential in the number of remaining indices. Graphically, the contraction of two tensors is represented by eliminating the edges (indexes) that the corresponding two vertices (tensors) have in common, and replacing both vertices with a single vertex with the remaining edges (indexes).  As in variable elimination, the cost also depends in the order in which indexes are eliminated.

Reference~\cite{markov_simulating_2008} relates the minimum cost of the tensor contraction to the treewidth of the corresponding line graph. The line graph $G'$ corresponding to the graph $G$ representing the tensor network is defined with a vertex for every edge in $G$. If two edges are incident in the same vertex (tensor) in $G$, then we add an edge in $G'$ between the corresponding two vertices of $G'$. In this way a two-qubit gate (a tensor with four indexes) results in a clique between four vertices in the line graph. For instance, the line graph corresponding to the tensor network in~\eqref{fig:ex_tn} is
\begin{align}
    \begin{tikzpicture}   
      \node (7) at (0,1) [circle,draw] {}; 
      \node (8) at (3,0) [circle,draw] {};
      \node (4) at (0,0) [circle,draw] {};
      \node (5) at (1,0) [circle,draw] {};
      \node (1) at (1,1) [circle,draw] {};
      \node (2) at (2,1) [circle,draw] {};
      \node (6) at (2,0) [circle,draw] {};
      \node (3) at (3,1) [circle,draw] {};
      \draw (4) to (5) to (6);
      \draw (5) to (1) to (2) to (3);
      \draw (6) to (2) to (5);
      \draw (6) to (1);
      \draw (7) to (1);
      \draw (6) to (8);
    \end{tikzpicture}\label{fig:ex_lg}\;.
  \end{align}

It follows that the line graph of the tensor network for a quantum
circuit is analogous to the undirected graphical model detailed
above. The main difference is that diagonal two-qubit (and
single-qubit) gates are treated more efficiently in the undirected
graphical model obtained directly from the quantum circuit factor
representation. In the line graph obtained from a tensor network a
diagonal gate results in a clique with four vertices, but in the
original graphical model is represented only by an edge and does not
introduce new vertices. This can be seen comparing the examples in Fig.~\ref{fig:ex_with_variables}, with treewidth 2, and~\eqref{fig:ex_lg}, which has treewidth 4. Because the cost is exponential in the treewidth, this optimization results in a smaller exponent for the computation.

\begin{figure}
  \centering
  \includegraphics[width=\columnwidth]{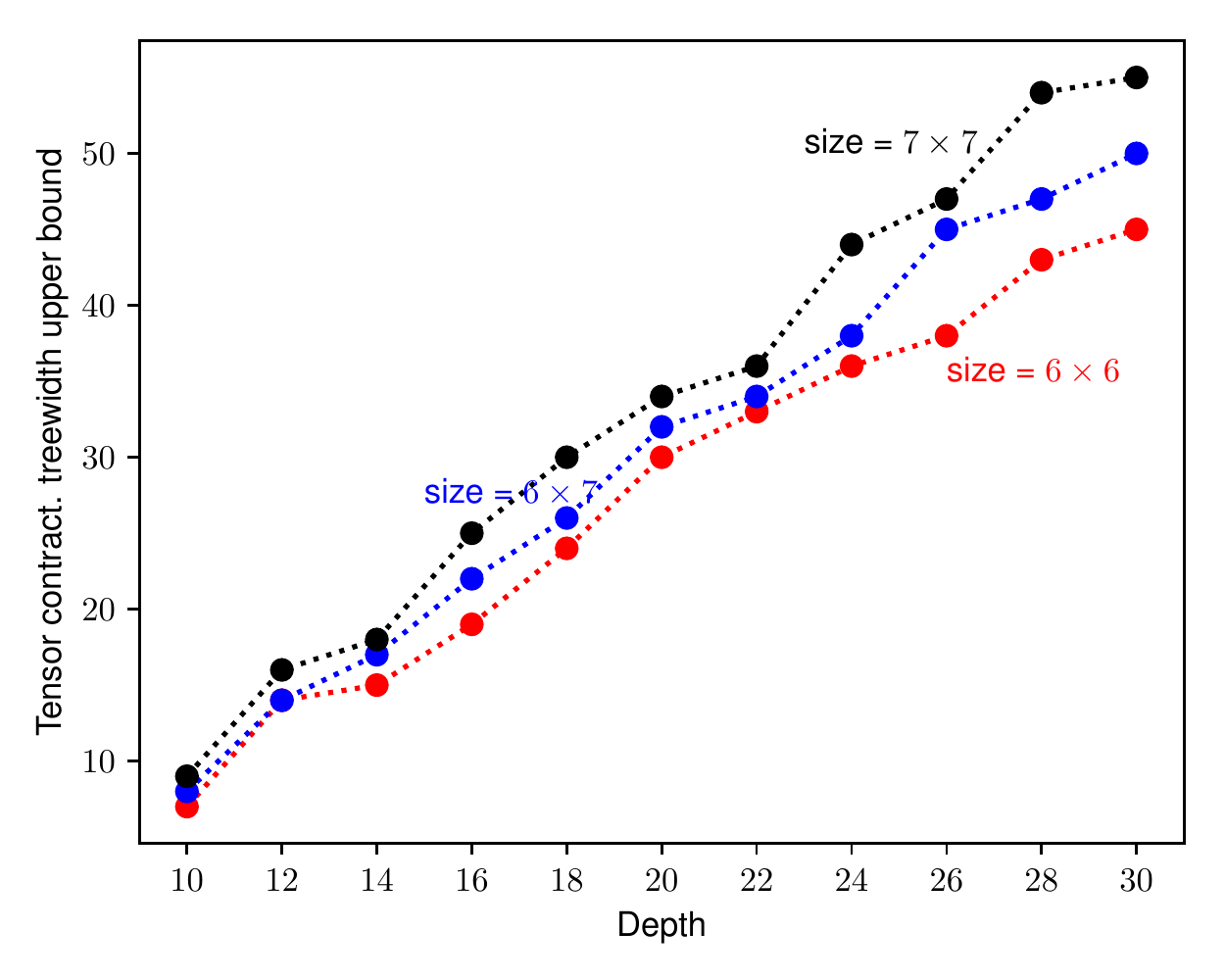}
  \caption{Numerical upper bound for the treewidth of the line graph which gives the tensor contraction complexity~\cite{markov_simulating_2008} of circuits from Ref.~\cite{boixo2016characterizing} with $6 \times 6$, $6 \times 7$, and $7 \times 7$ qubits as a function of the circuit depth.}
  \label{fig:treewidth_tensor}
\end{figure}

Figure~\ref{fig:treewidth_tensor} shows numerical upper bounds for the treewidth of the line graph as a function of depth for the circuits in Ref.~\cite{boixo2016characterizing}. The upper bounds were also obtained by running the \emph{QuickBB} algorithm~\cite{gogate2004complete}. Now that the treewidth has been reduced by almost a factor of two in the undirected graphical model which is drawn directly from the quantum circuit, Fig.~\ref{fig:treewidth}, instead of using the line graph of the tensor network, Fig.~\ref{fig:treewidth_tensor}. This is due to the fact that in these circuits all two-qubit gates are controlled-Z gates.

\section{The partition function for random circuits}\label{app:pf}
We review the direct mapping of random circuits to a partition
function of an Ising model at imaginary temperature from
Ref.~\cite{boixo2016characterizing}.  Previous algorithms for mapping
universal quantum circuits to partition functions of complex Ising
models use polynomial reductions to a universal gate
set~\cite{lidar_quantum_2004,geraci_classical_2010,de2011quantum},
which make them less efficient.

We encode each individual Feynman path by the set of $L$ binary
variables or spins $s = \{s_j^k\}$ with  $j\in[1\isep n]$, $k\in[0
\isep d_j]$ and $s_j^k \in \{1,-1\}$. One can easily see from  the  explicit form of the
non-diagonal gates in our example that  the absolute values of the probability amplitudes associated  with different  paths are all the same and equal to   $2^{-L/2}$. Using this fact we write the  path integral \eqref{eq:fpath} in the following form 
\begin{equation}
\langle x| \cU |0\ldots 0\rangle=2^{-L/2} \sum_{s}\exp\left(\frac{i\pi}{4} H_s(x) \right)\;.\label{eq:path1} 
\end{equation}
Here $\exp(i\pi H_s(x)/4)$ is  a phase factor associated with each path that depends explicitly on the end-point condition (\ref{eq:bc}).

The value of  the phase  $\pi H_s/4$ is accumulated as a sum of discrete phase changes that are associated with individual gates. For the $k$-th non-diagonal gate applied to qubit $j$ we introduce the coefficient $\alpha_j^k$ such that  $\alpha_j^k=1$ if the gate is  $\X^{1/2}$ and  $\alpha_j^k=0$ if the gate is $\Y^{1/2}$. 
 Thus, the  total phase change accumulated from the application of $\X^{1/2}$  and $\Y^{1/2}$  gates equals
\begin{align}
\hspace{-0.1in} \frac{i\pi}{4} H_{s}^{\X^{1/2}}(x) &=\frac{i \pi}{2} \sum_{j=1}^{n}\sum_{k=0}^{d_j}\alpha_j^k \frac{1+s_{j}^{k-1}s_{j}^{k}}{2}\;, \label{eq:phX} \\
\frac{i\pi}{4} H_{s}^{\Y^{1/2}}(x) &=i \pi  \sum_{j=1}^{n}\sum_{k=0}^{d_j}(1-\alpha_j^k) \frac{1-s_{j}^{k-1}}{2}\frac{1+s_j^k}{2}\;\nonumber\;.
\end{align}
As mentioned above, the dependence on $x$ arises due to the boundary condition (\ref{eq:bc}).  Note that we have omitted constant phase terms that do not depend on the path $s$.

We now describe the phase change from the action of  gates $\T$ and $\CZ$.  We introduce  coefficients $d(j,t)$  equal to the number of non-diagonal  gates applied to qubit $j$ over the first $t$ cycles (including the $0$-th cycle  of Hadamard gates).  We also introduce coefficients $\tau_j^t$ such that $\tau_j^t=1$ if a $\T$ gate is applied at cycle $t$ to qubit $j$ and $\tau_j^t=0$ otherwise.
Then the total  phase accumulated  from the action of the $\T$ gates equals
\begin{align}
\frac{i\pi}{4} H_{s}^{T}(x)=\frac{i \pi}{4} \sum_{j=1}^{n}\sum_{t=0}^{d}\tau_j^t \frac{1-s_{j}^{d(j,t)}}{2}\label{eq:phT}\;.
\end{align}
For a given pair of qubits $(i,j)$, we introduce coefficients $z_{ij}^{t}$ such that   $z_{ij}^{t}=1$ if  a $\CZ$ gate is applied to the  qubit pair during cycle $t$ and  $z_{ij}^{t}=0$ otherwise.
The total  phase accumulated  from the action of the $\CZ$ gates equals
\begin{multline}
\frac{ i\pi}{4} H_{s}^{\CZ}(x)\\=i \pi \sum_{i=1}^{n}\sum_{j=1}^{i-1}\sum_{t=0}^{d}z_{ij}^t  \frac{ 1-s_{i}^{d(i,t)}}{2}\frac{1-s_{j}^{d(j,t)}}{2}.\label{eq:phCZ}
\end{multline}

One can see from comparing   (\ref{eq:path1}) with (\ref{eq:phX})-(\ref{eq:phCZ}) that the wavefunction amplitudes $ \langle x| U |0\ldots 0\rangle$ take the form of a partition function of a classical Ising model 
with energy $H_s$ for a state $s$ and purely imaginary inverse
temperature $i \pi/4$. The total phase for each path  takes 8 distinct
values (mod 2$\pi$)  equal to $[0,\pi/4 \isep 7\pi/4]$. The
interaction graph of this Ising model is the undirected graphical
model introduced in Sec.~\ref{sec:ugm}.

Note that  if a quantum circuit uses only Clifford gates (not $\T$ gates), the total phase for each spin configuration in the partition function (mod 2$\pi$) is restricted to $[0,\pi/2,\pi,3\pi/2]$. In these case,  the corresponding partition function can be calculated efficiently in polynomial time~\cite{gottesman1998heisenberg,fujii2013quantum,goldberg_complexity_2014}. 

\newpage

\bibliographystyle{apsrev4-1} 
\bibliography{random_circuits}

\begin{thebibliography}{50}%
\makeatletter
\providecommand \@ifxundefined [1]{%
 \@ifx{#1\undefined}
}%
\providecommand \@ifnum [1]{%
 \ifnum #1\expandafter \@firstoftwo
 \else \expandafter \@secondoftwo
 \fi
}%
\providecommand \@ifx [1]{%
 \ifx #1\expandafter \@firstoftwo
 \else \expandafter \@secondoftwo
 \fi
}%
\providecommand \natexlab [1]{#1}%
\providecommand \enquote  [1]{``#1''}%
\providecommand \bibnamefont  [1]{#1}%
\providecommand \bibfnamefont [1]{#1}%
\providecommand \citenamefont [1]{#1}%
\providecommand \href@noop [0]{\@secondoftwo}%
\providecommand \href [0]{\begingroup \@sanitize@url \@href}%
\providecommand \@href[1]{\@@startlink{#1}\@@href}%
\providecommand \@@href[1]{\endgroup#1\@@endlink}%
\providecommand \@sanitize@url [0]{\catcode `\\12\catcode `\$12\catcode
  `\&12\catcode `\#12\catcode `\^12\catcode `\_12\catcode `\%12\relax}%
\providecommand \@@startlink[1]{}%
\providecommand \@@endlink[0]{}%
\providecommand \url  [0]{\begingroup\@sanitize@url \@url }%
\providecommand \@url [1]{\endgroup\@href {#1}{\urlprefix }}%
\providecommand \urlprefix  [0]{URL }%
\providecommand \Eprint [0]{\href }%
\providecommand \doibase [0]{http://dx.doi.org/}%
\providecommand \selectlanguage [0]{\@gobble}%
\providecommand \bibinfo  [0]{\@secondoftwo}%
\providecommand \bibfield  [0]{\@secondoftwo}%
\providecommand \translation [1]{[#1]}%
\providecommand \BibitemOpen [0]{}%
\providecommand \bibitemStop [0]{}%
\providecommand \bibitemNoStop [0]{.\EOS\space}%
\providecommand \EOS [0]{\spacefactor3000\relax}%
\providecommand \BibitemShut  [1]{\csname bibitem#1\endcsname}%
\let\auto@bib@innerbib\@empty
\bibitem [{\citenamefont {Boixo}\ \emph {et~al.}(2016)\citenamefont {Boixo},
  \citenamefont {Isakov}, \citenamefont {Smelyanskiy}, \citenamefont {Babbush},
  \citenamefont {Ding}, \citenamefont {Jiang}, \citenamefont {Martinis},\ and\
  \citenamefont {Neven}}]{boixo2016characterizing}%
  \BibitemOpen
  \bibfield  {author} {\bibinfo {author} {\bibfnamefont {S.}~\bibnamefont
  {Boixo}}, \bibinfo {author} {\bibfnamefont {S.~V.}\ \bibnamefont {Isakov}},
  \bibinfo {author} {\bibfnamefont {V.~N.}\ \bibnamefont {Smelyanskiy}},
  \bibinfo {author} {\bibfnamefont {R.}~\bibnamefont {Babbush}}, \bibinfo
  {author} {\bibfnamefont {N.}~\bibnamefont {Ding}}, \bibinfo {author}
  {\bibfnamefont {Z.}~\bibnamefont {Jiang}}, \bibinfo {author} {\bibfnamefont
  {J.~M.}\ \bibnamefont {Martinis}}, \ and\ \bibinfo {author} {\bibfnamefont
  {H.}~\bibnamefont {Neven}},\ }\href@noop {} {\bibfield  {journal} {\bibinfo
  {journal} {arXiv:1608.00263}\ } (\bibinfo {year} {2016})}\BibitemShut
  {NoStop}%
\bibitem [{\citenamefont {Feynman}(1982)}]{feynman1982simulating}%
  \BibitemOpen
  \bibfield  {author} {\bibinfo {author} {\bibfnamefont {R.~P.}\ \bibnamefont
  {Feynman}},\ }\href@noop {} {\bibfield  {journal} {\bibinfo  {journal} {Int.
  J. Theor. Phys.}\ }\textbf {\bibinfo {volume} {21}},\ \bibinfo {pages} {467}
  (\bibinfo {year} {1982})}\BibitemShut {NoStop}%
\bibitem [{\citenamefont {Aspuru-Guzik}\ \emph {et~al.}(2005)\citenamefont
  {Aspuru-Guzik}, \citenamefont {Dutoi}, \citenamefont {Love},\ and\
  \citenamefont {Head-Gordon}}]{aspuru2005simulated}%
  \BibitemOpen
  \bibfield  {author} {\bibinfo {author} {\bibfnamefont {A.}~\bibnamefont
  {Aspuru-Guzik}}, \bibinfo {author} {\bibfnamefont {A.~D.}\ \bibnamefont
  {Dutoi}}, \bibinfo {author} {\bibfnamefont {P.~J.}\ \bibnamefont {Love}}, \
  and\ \bibinfo {author} {\bibfnamefont {M.}~\bibnamefont {Head-Gordon}},\
  }\href@noop {} {\bibfield  {journal} {\bibinfo  {journal} {Science}\ }\textbf
  {\bibinfo {volume} {309}},\ \bibinfo {pages} {1704} (\bibinfo {year}
  {2005})}\BibitemShut {NoStop}%
\bibitem [{\citenamefont {Hastings}\ \emph {et~al.}(2015)\citenamefont
  {Hastings}, \citenamefont {Wecker}, \citenamefont {Bauer},\ and\
  \citenamefont {Troyer}}]{hastings2015improving}%
  \BibitemOpen
  \bibfield  {author} {\bibinfo {author} {\bibfnamefont {M.~B.}\ \bibnamefont
  {Hastings}}, \bibinfo {author} {\bibfnamefont {D.}~\bibnamefont {Wecker}},
  \bibinfo {author} {\bibfnamefont {B.}~\bibnamefont {Bauer}}, \ and\ \bibinfo
  {author} {\bibfnamefont {M.}~\bibnamefont {Troyer}},\ }\href@noop {}
  {\bibfield  {journal} {\bibinfo  {journal} {QIC}\ }\textbf {\bibinfo {volume}
  {15}},\ \bibinfo {pages} {1} (\bibinfo {year} {2015})}\BibitemShut {NoStop}%
\bibitem [{\citenamefont {Wecker}\ \emph {et~al.}(2015)\citenamefont {Wecker},
  \citenamefont {Hastings}, \citenamefont {Wiebe}, \citenamefont {Clark},
  \citenamefont {Nayak},\ and\ \citenamefont {Troyer}}]{wecker2015solving}%
  \BibitemOpen
  \bibfield  {author} {\bibinfo {author} {\bibfnamefont {D.}~\bibnamefont
  {Wecker}}, \bibinfo {author} {\bibfnamefont {M.~B.}\ \bibnamefont
  {Hastings}}, \bibinfo {author} {\bibfnamefont {N.}~\bibnamefont {Wiebe}},
  \bibinfo {author} {\bibfnamefont {B.~K.}\ \bibnamefont {Clark}}, \bibinfo
  {author} {\bibfnamefont {C.}~\bibnamefont {Nayak}}, \ and\ \bibinfo {author}
  {\bibfnamefont {M.}~\bibnamefont {Troyer}},\ }\href@noop {} {\bibfield
  {journal} {\bibinfo  {journal} {Phys. Rev. A}\ }\textbf {\bibinfo {volume}
  {92}},\ \bibinfo {pages} {062318} (\bibinfo {year} {2015})}\BibitemShut
  {NoStop}%
\bibitem [{\citenamefont {O’Malley}\ \emph {et~al.}(2016)\citenamefont
  {O’Malley}, \citenamefont {Babbush}, \citenamefont {Kivlichan},
  \citenamefont {Romero}, \citenamefont {McClean}, \citenamefont {Barends},
  \citenamefont {Kelly}, \citenamefont {Roushan}, \citenamefont {Tranter},
  \citenamefont {Ding} \emph {et~al.}}]{o2016scalable}%
  \BibitemOpen
  \bibfield  {author} {\bibinfo {author} {\bibfnamefont {P.}~\bibnamefont
  {O’Malley}}, \bibinfo {author} {\bibfnamefont {R.}~\bibnamefont {Babbush}},
  \bibinfo {author} {\bibfnamefont {I.}~\bibnamefont {Kivlichan}}, \bibinfo
  {author} {\bibfnamefont {J.}~\bibnamefont {Romero}}, \bibinfo {author}
  {\bibfnamefont {J.}~\bibnamefont {McClean}}, \bibinfo {author} {\bibfnamefont
  {R.}~\bibnamefont {Barends}}, \bibinfo {author} {\bibfnamefont
  {J.}~\bibnamefont {Kelly}}, \bibinfo {author} {\bibfnamefont
  {P.}~\bibnamefont {Roushan}}, \bibinfo {author} {\bibfnamefont
  {A.}~\bibnamefont {Tranter}}, \bibinfo {author} {\bibfnamefont
  {N.}~\bibnamefont {Ding}},  \emph {et~al.},\ }\href@noop {} {\bibfield
  {journal} {\bibinfo  {journal} {Phys. Rev. X}\ }\textbf {\bibinfo {volume}
  {6}},\ \bibinfo {pages} {031007} (\bibinfo {year} {2016})}\BibitemShut
  {NoStop}%
\bibitem [{\citenamefont {Reiher}\ \emph {et~al.}(2017)\citenamefont {Reiher},
  \citenamefont {Wiebe}, \citenamefont {Svore}, \citenamefont {Wecker},\ and\
  \citenamefont {Troyer}}]{reiher2017elucidating}%
  \BibitemOpen
  \bibfield  {author} {\bibinfo {author} {\bibfnamefont {M.}~\bibnamefont
  {Reiher}}, \bibinfo {author} {\bibfnamefont {N.}~\bibnamefont {Wiebe}},
  \bibinfo {author} {\bibfnamefont {K.~M.}\ \bibnamefont {Svore}}, \bibinfo
  {author} {\bibfnamefont {D.}~\bibnamefont {Wecker}}, \ and\ \bibinfo {author}
  {\bibfnamefont {M.}~\bibnamefont {Troyer}},\ }\href@noop {} {\bibfield
  {journal} {\bibinfo  {journal} {PNAS}\ ,\ \bibinfo {pages} {201619152}}
  (\bibinfo {year} {2017})}\BibitemShut {NoStop}%
\bibitem [{\citenamefont {Babbush}\ \emph {et~al.}(2017)\citenamefont
  {Babbush}, \citenamefont {Wiebe}, \citenamefont {McClean}, \citenamefont
  {McClain}, \citenamefont {Neven},\ and\ \citenamefont
  {Chan}}]{babbush2017low}%
  \BibitemOpen
  \bibfield  {author} {\bibinfo {author} {\bibfnamefont {R.}~\bibnamefont
  {Babbush}}, \bibinfo {author} {\bibfnamefont {N.}~\bibnamefont {Wiebe}},
  \bibinfo {author} {\bibfnamefont {J.}~\bibnamefont {McClean}}, \bibinfo
  {author} {\bibfnamefont {J.}~\bibnamefont {McClain}}, \bibinfo {author}
  {\bibfnamefont {H.}~\bibnamefont {Neven}}, \ and\ \bibinfo {author}
  {\bibfnamefont {G.~K.}\ \bibnamefont {Chan}},\ }\href@noop {} {\bibfield
  {journal} {\bibinfo  {journal} {arXiv:1706.00023}\ } (\bibinfo {year}
  {2017})}\BibitemShut {NoStop}%
\bibitem [{\citenamefont {Jiang}\ \emph {et~al.}(2017)\citenamefont {Jiang},
  \citenamefont {Sung}, \citenamefont {Kechedzhi}, \citenamefont
  {Smelyanskiy},\ and\ \citenamefont {Boixo}}]{jiang2017quantum}%
  \BibitemOpen
  \bibfield  {author} {\bibinfo {author} {\bibfnamefont {Z.}~\bibnamefont
  {Jiang}}, \bibinfo {author} {\bibfnamefont {K.~J.}\ \bibnamefont {Sung}},
  \bibinfo {author} {\bibfnamefont {K.}~\bibnamefont {Kechedzhi}}, \bibinfo
  {author} {\bibfnamefont {V.~N.}\ \bibnamefont {Smelyanskiy}}, \ and\ \bibinfo
  {author} {\bibfnamefont {S.}~\bibnamefont {Boixo}},\ }\href@noop {}
  {\bibfield  {journal} {\bibinfo  {journal} {arXiv:1711.05395}\ } (\bibinfo
  {year} {2017})}\BibitemShut {NoStop}%
\bibitem [{\citenamefont {Shor}(1999)}]{shor1999polynomial}%
  \BibitemOpen
  \bibfield  {author} {\bibinfo {author} {\bibfnamefont {P.~W.}\ \bibnamefont
  {Shor}},\ }\href@noop {} {\bibfield  {journal} {\bibinfo  {journal} {SIAM
  review}\ }\textbf {\bibinfo {volume} {41}},\ \bibinfo {pages} {303} (\bibinfo
  {year} {1999})}\BibitemShut {NoStop}%
\bibitem [{\citenamefont {Aaronson}\ and\ \citenamefont
  {Arkhipov}(2011)}]{aaronson2011computational}%
  \BibitemOpen
  \bibfield  {author} {\bibinfo {author} {\bibfnamefont {S.}~\bibnamefont
  {Aaronson}}\ and\ \bibinfo {author} {\bibfnamefont {A.}~\bibnamefont
  {Arkhipov}},\ }in\ \href@noop {} {\emph {\bibinfo {booktitle} {STOC}}}\
  (\bibinfo {organization} {ACM},\ \bibinfo {year} {2011})\ pp.\ \bibinfo
  {pages} {333--342}\BibitemShut {NoStop}%
\bibitem [{\citenamefont {Bremner}\ \emph
  {et~al.}(2016{\natexlab{a}})\citenamefont {Bremner}, \citenamefont
  {Montanaro},\ and\ \citenamefont {Shepherd}}]{bremner2015average}%
  \BibitemOpen
  \bibfield  {author} {\bibinfo {author} {\bibfnamefont {M.~J.}\ \bibnamefont
  {Bremner}}, \bibinfo {author} {\bibfnamefont {A.}~\bibnamefont {Montanaro}},
  \ and\ \bibinfo {author} {\bibfnamefont {D.~J.}\ \bibnamefont {Shepherd}},\
  }\href@noop {} {\bibfield  {journal} {\bibinfo  {journal} {Phys. Rev. Lett.}\
  }\textbf {\bibinfo {volume} {117}},\ \bibinfo {pages} {080501} (\bibinfo
  {year} {2016}{\natexlab{a}})}\BibitemShut {NoStop}%
\bibitem [{\citenamefont {Bremner}\ \emph
  {et~al.}(2016{\natexlab{b}})\citenamefont {Bremner}, \citenamefont
  {Montanaro},\ and\ \citenamefont {Shepherd}}]{bremner16}%
  \BibitemOpen
  \bibfield  {author} {\bibinfo {author} {\bibfnamefont {M.~J.}\ \bibnamefont
  {Bremner}}, \bibinfo {author} {\bibfnamefont {A.}~\bibnamefont {Montanaro}},
  \ and\ \bibinfo {author} {\bibfnamefont {D.~J.}\ \bibnamefont {Shepherd}},\
  }\href@noop {} {\bibfield  {journal} {\bibinfo  {journal} {arXiv:1610.01808}\
  } (\bibinfo {year} {2016}{\natexlab{b}})}\BibitemShut {NoStop}%
\bibitem [{\citenamefont {Gao}\ \emph {et~al.}(2017)\citenamefont {Gao},
  \citenamefont {Wang},\ and\ \citenamefont {Duan}}]{gao2017quantum}%
  \BibitemOpen
  \bibfield  {author} {\bibinfo {author} {\bibfnamefont {X.}~\bibnamefont
  {Gao}}, \bibinfo {author} {\bibfnamefont {S.-T.}\ \bibnamefont {Wang}}, \
  and\ \bibinfo {author} {\bibfnamefont {L.-M.}\ \bibnamefont {Duan}},\
  }\href@noop {} {\bibfield  {journal} {\bibinfo  {journal} {Physical Review
  Letters}\ }\textbf {\bibinfo {volume} {118}},\ \bibinfo {pages} {040502}
  (\bibinfo {year} {2017})}\BibitemShut {NoStop}%
\bibitem [{\citenamefont {Hangleiter}\ \emph {et~al.}(2017)\citenamefont
  {Hangleiter}, \citenamefont {Kliesch}, \citenamefont {Schwarz},\ and\
  \citenamefont {Eisert}}]{hangleiter2017direct}%
  \BibitemOpen
  \bibfield  {author} {\bibinfo {author} {\bibfnamefont {D.}~\bibnamefont
  {Hangleiter}}, \bibinfo {author} {\bibfnamefont {M.}~\bibnamefont {Kliesch}},
  \bibinfo {author} {\bibfnamefont {M.}~\bibnamefont {Schwarz}}, \ and\
  \bibinfo {author} {\bibfnamefont {J.}~\bibnamefont {Eisert}},\ }\href@noop {}
  {\bibfield  {journal} {\bibinfo  {journal} {QST}\ }\textbf {\bibinfo {volume}
  {2}},\ \bibinfo {pages} {015004} (\bibinfo {year} {2017})}\BibitemShut
  {NoStop}%
\bibitem [{\citenamefont {Kapourniotis}\ and\ \citenamefont
  {Datta}(2017)}]{kapourniotis2017nonadaptive}%
  \BibitemOpen
  \bibfield  {author} {\bibinfo {author} {\bibfnamefont {T.}~\bibnamefont
  {Kapourniotis}}\ and\ \bibinfo {author} {\bibfnamefont {A.}~\bibnamefont
  {Datta}},\ }\href@noop {} {\bibfield  {journal} {\bibinfo  {journal}
  {arXiv:1703.09568}\ } (\bibinfo {year} {2017})}\BibitemShut {NoStop}%
\bibitem [{\citenamefont {Neill}\ \emph {et~al.}(2017)\citenamefont {Neill},
  \citenamefont {Roushan}, \citenamefont {Kechedzhi}, \citenamefont {Boixo},
  \citenamefont {Isakov}, \citenamefont {Smelyanskiy}, \citenamefont {Barends},
  \citenamefont {Burkett}, \citenamefont {Chen}, \citenamefont {Chen},
  \citenamefont {Chiaro}, \citenamefont {Dunsworth}, \citenamefont {Fowler},
  \citenamefont {Foxen}, \citenamefont {Graff}, \citenamefont {Jeffrey},
  \citenamefont {Kelly}, \citenamefont {Lucero}, \citenamefont {Megrant},
  \citenamefont {Mutus}, \citenamefont {Neeley}, \citenamefont {Quintana},
  \citenamefont {Sank}, \citenamefont {Vainsencher}, \citenamefont {Wenner},
  \citenamefont {White}, \citenamefont {Neven},\ and\ \citenamefont
  {Martinis}}]{neill2017}%
  \BibitemOpen
  \bibfield  {author} {\bibinfo {author} {\bibfnamefont {C.}~\bibnamefont
  {Neill}}, \bibinfo {author} {\bibfnamefont {P.}~\bibnamefont {Roushan}},
  \bibinfo {author} {\bibfnamefont {K.}~\bibnamefont {Kechedzhi}}, \bibinfo
  {author} {\bibfnamefont {S.}~\bibnamefont {Boixo}}, \bibinfo {author}
  {\bibfnamefont {S.~V.}\ \bibnamefont {Isakov}}, \bibinfo {author}
  {\bibfnamefont {V.}~\bibnamefont {Smelyanskiy}}, \bibinfo {author}
  {\bibfnamefont {R.}~\bibnamefont {Barends}}, \bibinfo {author} {\bibfnamefont
  {B.}~\bibnamefont {Burkett}}, \bibinfo {author} {\bibfnamefont
  {Y.}~\bibnamefont {Chen}}, \bibinfo {author} {\bibfnamefont {Z.}~\bibnamefont
  {Chen}}, \bibinfo {author} {\bibfnamefont {B.}~\bibnamefont {Chiaro}},
  \bibinfo {author} {\bibfnamefont {A.}~\bibnamefont {Dunsworth}}, \bibinfo
  {author} {\bibfnamefont {A.}~\bibnamefont {Fowler}}, \bibinfo {author}
  {\bibfnamefont {B.}~\bibnamefont {Foxen}}, \bibinfo {author} {\bibfnamefont
  {R.}~\bibnamefont {Graff}}, \bibinfo {author} {\bibfnamefont
  {E.}~\bibnamefont {Jeffrey}}, \bibinfo {author} {\bibfnamefont
  {J.}~\bibnamefont {Kelly}}, \bibinfo {author} {\bibfnamefont
  {E.}~\bibnamefont {Lucero}}, \bibinfo {author} {\bibfnamefont
  {A.}~\bibnamefont {Megrant}}, \bibinfo {author} {\bibfnamefont
  {J.}~\bibnamefont {Mutus}}, \bibinfo {author} {\bibfnamefont
  {M.}~\bibnamefont {Neeley}}, \bibinfo {author} {\bibfnamefont
  {C.}~\bibnamefont {Quintana}}, \bibinfo {author} {\bibfnamefont
  {D.}~\bibnamefont {Sank}}, \bibinfo {author} {\bibfnamefont {A.}~\bibnamefont
  {Vainsencher}}, \bibinfo {author} {\bibfnamefont {J.}~\bibnamefont {Wenner}},
  \bibinfo {author} {\bibfnamefont {T.~C.}\ \bibnamefont {White}}, \bibinfo
  {author} {\bibfnamefont {H.}~\bibnamefont {Neven}}, \ and\ \bibinfo {author}
  {\bibfnamefont {J.~M.}\ \bibnamefont {Martinis}},\ }\href@noop {} {\bibfield
  {journal} {\bibinfo  {journal} {arXiv:1709.06678}\ } (\bibinfo {year}
  {2017})}\BibitemShut {NoStop}%
\bibitem [{\citenamefont {Schack}\ and\ \citenamefont
  {Caves}(1993)}]{schack_hypersensitivity_1993}%
  \BibitemOpen
  \bibfield  {author} {\bibinfo {author} {\bibfnamefont {R.}~\bibnamefont
  {Schack}}\ and\ \bibinfo {author} {\bibfnamefont {C.~M.}\ \bibnamefont
  {Caves}},\ }\href@noop {} {\bibfield  {journal} {\bibinfo  {journal} {Phys.
  Rev. Lett.}\ }\textbf {\bibinfo {volume} {71}},\ \bibinfo {pages} {525}
  (\bibinfo {year} {1993})}\BibitemShut {NoStop}%
\bibitem [{\citenamefont {Scott}\ \emph {et~al.}(2006)\citenamefont {Scott},
  \citenamefont {Brun}, \citenamefont {Caves},\ and\ \citenamefont
  {Schack}}]{scott_hypersensitivity_2006}%
  \BibitemOpen
  \bibfield  {author} {\bibinfo {author} {\bibfnamefont {A.~J.}\ \bibnamefont
  {Scott}}, \bibinfo {author} {\bibfnamefont {T.~A.}\ \bibnamefont {Brun}},
  \bibinfo {author} {\bibfnamefont {C.~M.}\ \bibnamefont {Caves}}, \ and\
  \bibinfo {author} {\bibfnamefont {R.}~\bibnamefont {Schack}},\ }\href@noop {}
  {\bibfield  {journal} {\bibinfo  {journal} {J. Phys. A: Math. Gen.}\ }\textbf
  {\bibinfo {volume} {39}},\ \bibinfo {pages} {13405} (\bibinfo {year}
  {2006})}\BibitemShut {NoStop}%
\bibitem [{\citenamefont {Barends}\ \emph {et~al.}(2014)\citenamefont
  {Barends}, \citenamefont {Kelly}, \citenamefont {Megrant}, \citenamefont
  {Veitia}, \citenamefont {Sank}, \citenamefont {Jeffrey}, \citenamefont
  {White}, \citenamefont {Mutus}, \citenamefont {Fowler}, \citenamefont
  {Campbell},\ and\ \citenamefont {{others}}}]{barends_superconducting_2014}%
  \BibitemOpen
  \bibfield  {author} {\bibinfo {author} {\bibfnamefont {R.}~\bibnamefont
  {Barends}}, \bibinfo {author} {\bibfnamefont {J.}~\bibnamefont {Kelly}},
  \bibinfo {author} {\bibfnamefont {A.}~\bibnamefont {Megrant}}, \bibinfo
  {author} {\bibfnamefont {A.}~\bibnamefont {Veitia}}, \bibinfo {author}
  {\bibfnamefont {D.}~\bibnamefont {Sank}}, \bibinfo {author} {\bibfnamefont
  {E.}~\bibnamefont {Jeffrey}}, \bibinfo {author} {\bibfnamefont {T.~C.}\
  \bibnamefont {White}}, \bibinfo {author} {\bibfnamefont {J.}~\bibnamefont
  {Mutus}}, \bibinfo {author} {\bibfnamefont {A.~G.}\ \bibnamefont {Fowler}},
  \bibinfo {author} {\bibfnamefont {B.}~\bibnamefont {Campbell}}, \ and\
  \bibinfo {author} {\bibnamefont {{others}}},\ }\href
  {http://www.nature.com/nature/journal/v508/n7497/abs/nature13171.html}
  {\bibfield  {journal} {\bibinfo  {journal} {Nature}\ }\textbf {\bibinfo
  {volume} {508}},\ \bibinfo {pages} {500} (\bibinfo {year}
  {2014})}\BibitemShut {NoStop}%
\bibitem [{\citenamefont {Fowler}\ \emph {et~al.}(2012)\citenamefont {Fowler},
  \citenamefont {Mariantoni}, \citenamefont {Martinis},\ and\ \citenamefont
  {Cleland}}]{fowler2012surface}%
  \BibitemOpen
  \bibfield  {author} {\bibinfo {author} {\bibfnamefont {A.~G.}\ \bibnamefont
  {Fowler}}, \bibinfo {author} {\bibfnamefont {M.}~\bibnamefont {Mariantoni}},
  \bibinfo {author} {\bibfnamefont {J.~M.}\ \bibnamefont {Martinis}}, \ and\
  \bibinfo {author} {\bibfnamefont {A.~N.}\ \bibnamefont {Cleland}},\
  }\href@noop {} {\bibfield  {journal} {\bibinfo  {journal} {Phys. Rev. A}\
  }\textbf {\bibinfo {volume} {86}},\ \bibinfo {pages} {032324} (\bibinfo
  {year} {2012})}\BibitemShut {NoStop}%
\bibitem [{\citenamefont {De~Raedt}\ \emph {et~al.}(2007)\citenamefont
  {De~Raedt}, \citenamefont {Michielsen}, \citenamefont {De~Raedt},
  \citenamefont {Trieu}, \citenamefont {Arnold}, \citenamefont {Richter},
  \citenamefont {Lippert}, \citenamefont {Watanabe},\ and\ \citenamefont
  {Ito}}]{de2007massively}%
  \BibitemOpen
  \bibfield  {author} {\bibinfo {author} {\bibfnamefont {K.}~\bibnamefont
  {De~Raedt}}, \bibinfo {author} {\bibfnamefont {K.}~\bibnamefont
  {Michielsen}}, \bibinfo {author} {\bibfnamefont {H.}~\bibnamefont
  {De~Raedt}}, \bibinfo {author} {\bibfnamefont {B.}~\bibnamefont {Trieu}},
  \bibinfo {author} {\bibfnamefont {G.}~\bibnamefont {Arnold}}, \bibinfo
  {author} {\bibfnamefont {M.}~\bibnamefont {Richter}}, \bibinfo {author}
  {\bibfnamefont {T.}~\bibnamefont {Lippert}}, \bibinfo {author} {\bibfnamefont
  {H.}~\bibnamefont {Watanabe}}, \ and\ \bibinfo {author} {\bibfnamefont
  {N.}~\bibnamefont {Ito}},\ }\href@noop {} {\bibfield  {journal} {\bibinfo
  {journal} {Computer Physics Communications}\ }\textbf {\bibinfo {volume}
  {176}},\ \bibinfo {pages} {121} (\bibinfo {year} {2007})}\BibitemShut
  {NoStop}%
\bibitem [{\citenamefont {Zulehner}\ and\ \citenamefont
  {Wille}(2017)}]{zulehner2017advanced}%
  \BibitemOpen
  \bibfield  {author} {\bibinfo {author} {\bibfnamefont {A.}~\bibnamefont
  {Zulehner}}\ and\ \bibinfo {author} {\bibfnamefont {R.}~\bibnamefont
  {Wille}},\ }\href@noop {} {\bibfield  {journal} {\bibinfo  {journal} {arXiv
  preprint arXiv:1707.00865}\ } (\bibinfo {year} {2017})}\BibitemShut {NoStop}%
\bibitem [{\citenamefont {H{\"a}ner}\ and\ \citenamefont
  {Steiger}(2017)}]{haner2017petabyte}%
  \BibitemOpen
  \bibfield  {author} {\bibinfo {author} {\bibfnamefont {T.}~\bibnamefont
  {H{\"a}ner}}\ and\ \bibinfo {author} {\bibfnamefont {D.~S.}\ \bibnamefont
  {Steiger}},\ }\href@noop {} {\bibfield  {journal} {\bibinfo  {journal}
  {arXiv:1704.01127}\ } (\bibinfo {year} {2017})}\BibitemShut {NoStop}%
\bibitem [{\citenamefont {Pednault}\ \emph {et~al.}(2017)\citenamefont
  {Pednault}, \citenamefont {Gunnels}, \citenamefont {Nannicini}, \citenamefont
  {Horesh}, \citenamefont {Magerlein}, \citenamefont {Solomonik},\ and\
  \citenamefont {Wisnieff}}]{pednault2017breaking}%
  \BibitemOpen
  \bibfield  {author} {\bibinfo {author} {\bibfnamefont {E.}~\bibnamefont
  {Pednault}}, \bibinfo {author} {\bibfnamefont {J.~A.}\ \bibnamefont
  {Gunnels}}, \bibinfo {author} {\bibfnamefont {G.}~\bibnamefont {Nannicini}},
  \bibinfo {author} {\bibfnamefont {L.}~\bibnamefont {Horesh}}, \bibinfo
  {author} {\bibfnamefont {T.}~\bibnamefont {Magerlein}}, \bibinfo {author}
  {\bibfnamefont {E.}~\bibnamefont {Solomonik}}, \ and\ \bibinfo {author}
  {\bibfnamefont {R.}~\bibnamefont {Wisnieff}},\ }\href@noop {} {\bibfield
  {journal} {\bibinfo  {journal} {arXiv:1710.05867}\ } (\bibinfo {year}
  {2017})}\BibitemShut {NoStop}%
\bibitem [{\citenamefont {Viamontes}\ \emph {et~al.}(2009)\citenamefont
  {Viamontes}, \citenamefont {Markov},\ and\ \citenamefont
  {Hayes}}]{viamontes2009quantum}%
  \BibitemOpen
  \bibfield  {author} {\bibinfo {author} {\bibfnamefont {G.~F.}\ \bibnamefont
  {Viamontes}}, \bibinfo {author} {\bibfnamefont {I.~L.}\ \bibnamefont
  {Markov}}, \ and\ \bibinfo {author} {\bibfnamefont {J.~P.}\ \bibnamefont
  {Hayes}},\ }\href@noop {} {\emph {\bibinfo {title} {Quantum circuit
  simulation}}}\ (\bibinfo  {publisher} {Springer Science \& Business Media},\
  \bibinfo {year} {2009})\BibitemShut {NoStop}%
\bibitem [{\citenamefont {Dechter}(1998)}]{dechter1998bucket}%
  \BibitemOpen
  \bibfield  {author} {\bibinfo {author} {\bibfnamefont {R.}~\bibnamefont
  {Dechter}},\ }in\ \href@noop {} {\emph {\bibinfo {booktitle} {Learning in
  graphical models}}}\ (\bibinfo  {publisher} {Springer},\ \bibinfo {year}
  {1998})\ pp.\ \bibinfo {pages} {75--104}\BibitemShut {NoStop}%
\bibitem [{\citenamefont {Murphy}(2012)}]{murphy_machine_2012}%
  \BibitemOpen
  \bibfield  {author} {\bibinfo {author} {\bibfnamefont {K.~P.}\ \bibnamefont
  {Murphy}},\ }\href@noop {} {\emph {\bibinfo {title} {Machine learning: a
  probabilistic perspective}}},\ Adaptive computation and machine learning
  series\ (\bibinfo  {publisher} {MIT Press},\ \bibinfo {address} {Cambridge,
  MA},\ \bibinfo {year} {2012})\BibitemShut {NoStop}%
\bibitem [{\citenamefont {Bernstein}\ and\ \citenamefont
  {Vazirani}(1997)}]{bernstein1997quantum}%
  \BibitemOpen
  \bibfield  {author} {\bibinfo {author} {\bibfnamefont {E.}~\bibnamefont
  {Bernstein}}\ and\ \bibinfo {author} {\bibfnamefont {U.}~\bibnamefont
  {Vazirani}},\ }\href@noop {} {\bibfield  {journal} {\bibinfo  {journal}
  {SIAMCOMP}\ }\textbf {\bibinfo {volume} {26}},\ \bibinfo {pages} {1411}
  (\bibinfo {year} {1997})}\BibitemShut {NoStop}%
\bibitem [{\citenamefont {Markov}\ and\ \citenamefont
  {Shi}(2008)}]{markov_simulating_2008}%
  \BibitemOpen
  \bibfield  {author} {\bibinfo {author} {\bibfnamefont {I.~L.}\ \bibnamefont
  {Markov}}\ and\ \bibinfo {author} {\bibfnamefont {Y.}~\bibnamefont {Shi}},\
  }\href {\doibase 10.1137/050644756} {\bibfield  {journal} {\bibinfo
  {journal} {SICOMP}\ }\textbf {\bibinfo {volume} {38}},\ \bibinfo {pages}
  {963} (\bibinfo {year} {2008})}\BibitemShut {NoStop}%
\bibitem [{\citenamefont {Aaronson}\ and\ \citenamefont
  {Chen}(2016)}]{aaronson2016complexity}%
  \BibitemOpen
  \bibfield  {author} {\bibinfo {author} {\bibfnamefont {S.}~\bibnamefont
  {Aaronson}}\ and\ \bibinfo {author} {\bibfnamefont {L.}~\bibnamefont
  {Chen}},\ }\href@noop {} {\bibfield  {journal} {\bibinfo  {journal}
  {arXiv:1612.05903}\ } (\bibinfo {year} {2016})}\BibitemShut {NoStop}%
\bibitem [{\citenamefont {Martinis}\ and\ \citenamefont
  {Geller}(2014)}]{martinis_fast_2014}%
  \BibitemOpen
  \bibfield  {author} {\bibinfo {author} {\bibfnamefont {J.~M.}\ \bibnamefont
  {Martinis}}\ and\ \bibinfo {author} {\bibfnamefont {M.~R.}\ \bibnamefont
  {Geller}},\ }\href {\doibase 10.1103/PhysRevA.90.022307} {\bibfield
  {journal} {\bibinfo  {journal} {Physical Review A}\ }\textbf {\bibinfo
  {volume} {90}} (\bibinfo {year} {2014}),\
  10.1103/PhysRevA.90.022307}\BibitemShut {NoStop}%
\bibitem [{\citenamefont {Barends}\ \emph {et~al.}(2015)\citenamefont
  {Barends}, \citenamefont {Lamata}, \citenamefont {Kelly}, \citenamefont
  {García-Álvarez}, \citenamefont {Fowler}, \citenamefont {Megrant},
  \citenamefont {Jeffrey}, \citenamefont {White}, \citenamefont {Sank},
  \citenamefont {Mutus}, \citenamefont {Campbell}, \citenamefont {Chen},
  \citenamefont {Chen}, \citenamefont {Chiaro}, \citenamefont {Dunsworth},
  \citenamefont {Hoi}, \citenamefont {Neill}, \citenamefont {O’Malley},
  \citenamefont {Quintana}, \citenamefont {Roushan}, \citenamefont
  {Vainsencher}, \citenamefont {Wenner}, \citenamefont {Solano},\ and\
  \citenamefont {Martinis}}]{barends_digital_2015}%
  \BibitemOpen
  \bibfield  {author} {\bibinfo {author} {\bibfnamefont {R.}~\bibnamefont
  {Barends}}, \bibinfo {author} {\bibfnamefont {L.}~\bibnamefont {Lamata}},
  \bibinfo {author} {\bibfnamefont {J.}~\bibnamefont {Kelly}}, \bibinfo
  {author} {\bibfnamefont {L.}~\bibnamefont {García-Álvarez}}, \bibinfo
  {author} {\bibfnamefont {A.~G.}\ \bibnamefont {Fowler}}, \bibinfo {author}
  {\bibfnamefont {A.}~\bibnamefont {Megrant}}, \bibinfo {author} {\bibfnamefont
  {E.}~\bibnamefont {Jeffrey}}, \bibinfo {author} {\bibfnamefont {T.~C.}\
  \bibnamefont {White}}, \bibinfo {author} {\bibfnamefont {D.}~\bibnamefont
  {Sank}}, \bibinfo {author} {\bibfnamefont {J.~Y.}\ \bibnamefont {Mutus}},
  \bibinfo {author} {\bibfnamefont {B.}~\bibnamefont {Campbell}}, \bibinfo
  {author} {\bibfnamefont {Y.}~\bibnamefont {Chen}}, \bibinfo {author}
  {\bibfnamefont {Z.}~\bibnamefont {Chen}}, \bibinfo {author} {\bibfnamefont
  {B.}~\bibnamefont {Chiaro}}, \bibinfo {author} {\bibfnamefont
  {A.}~\bibnamefont {Dunsworth}}, \bibinfo {author} {\bibfnamefont {I.-C.}\
  \bibnamefont {Hoi}}, \bibinfo {author} {\bibfnamefont {C.}~\bibnamefont
  {Neill}}, \bibinfo {author} {\bibfnamefont {P.~J.~J.}\ \bibnamefont
  {O’Malley}}, \bibinfo {author} {\bibfnamefont {C.}~\bibnamefont
  {Quintana}}, \bibinfo {author} {\bibfnamefont {P.}~\bibnamefont {Roushan}},
  \bibinfo {author} {\bibfnamefont {A.}~\bibnamefont {Vainsencher}}, \bibinfo
  {author} {\bibfnamefont {J.}~\bibnamefont {Wenner}}, \bibinfo {author}
  {\bibfnamefont {E.}~\bibnamefont {Solano}}, \ and\ \bibinfo {author}
  {\bibfnamefont {J.~M.}\ \bibnamefont {Martinis}},\ }\href
  {http://www.nature.com/doifinder/10.1038/ncomms8654} {\bibfield  {journal}
  {\bibinfo  {journal} {Nat. Comm.}\ }\textbf {\bibinfo {volume} {6}},\
  \bibinfo {pages} {7654} (\bibinfo {year} {2015})}\BibitemShut {NoStop}%
\bibitem [{\citenamefont {Kelly}\ \emph {et~al.}(2015)\citenamefont {Kelly},
  \citenamefont {Barends}, \citenamefont {Fowler}, \citenamefont {Megrant},
  \citenamefont {Jeffrey}, \citenamefont {White}, \citenamefont {Sank},
  \citenamefont {Mutus}, \citenamefont {Campbell}, \citenamefont {Chen},\ and\
  \citenamefont {{others}}}]{kelly_state_2015}%
  \BibitemOpen
  \bibfield  {author} {\bibinfo {author} {\bibfnamefont {J.}~\bibnamefont
  {Kelly}}, \bibinfo {author} {\bibfnamefont {R.}~\bibnamefont {Barends}},
  \bibinfo {author} {\bibfnamefont {A.~G.}\ \bibnamefont {Fowler}}, \bibinfo
  {author} {\bibfnamefont {A.}~\bibnamefont {Megrant}}, \bibinfo {author}
  {\bibfnamefont {E.}~\bibnamefont {Jeffrey}}, \bibinfo {author} {\bibfnamefont
  {T.~C.}\ \bibnamefont {White}}, \bibinfo {author} {\bibfnamefont
  {D.}~\bibnamefont {Sank}}, \bibinfo {author} {\bibfnamefont {J.~Y.}\
  \bibnamefont {Mutus}}, \bibinfo {author} {\bibfnamefont {B.}~\bibnamefont
  {Campbell}}, \bibinfo {author} {\bibfnamefont {Y.}~\bibnamefont {Chen}}, \
  and\ \bibinfo {author} {\bibnamefont {{others}}},\ }\href
  {http://www.nature.com/nature/journal/v519/n7541/abs/nature14270.html}
  {\bibfield  {journal} {\bibinfo  {journal} {Nature}\ }\textbf {\bibinfo
  {volume} {519}},\ \bibinfo {pages} {66} (\bibinfo {year} {2015})}\BibitemShut
  {NoStop}%
\bibitem [{\citenamefont {Barends}\ \emph {et~al.}(2016)\citenamefont
  {Barends}, \citenamefont {Shabani}, \citenamefont {Lamata}, \citenamefont
  {Kelly}, \citenamefont {Mezzacapo}, \citenamefont {Las~Heras}, \citenamefont
  {Babbush}, \citenamefont {Fowler}, \citenamefont {Campbell}, \citenamefont
  {Chen} \emph {et~al.}}]{barends_digitized_2015}%
  \BibitemOpen
  \bibfield  {author} {\bibinfo {author} {\bibfnamefont {R.}~\bibnamefont
  {Barends}}, \bibinfo {author} {\bibfnamefont {A.}~\bibnamefont {Shabani}},
  \bibinfo {author} {\bibfnamefont {L.}~\bibnamefont {Lamata}}, \bibinfo
  {author} {\bibfnamefont {J.}~\bibnamefont {Kelly}}, \bibinfo {author}
  {\bibfnamefont {A.}~\bibnamefont {Mezzacapo}}, \bibinfo {author}
  {\bibfnamefont {U.}~\bibnamefont {Las~Heras}}, \bibinfo {author}
  {\bibfnamefont {R.}~\bibnamefont {Babbush}}, \bibinfo {author} {\bibfnamefont
  {A.}~\bibnamefont {Fowler}}, \bibinfo {author} {\bibfnamefont
  {B.}~\bibnamefont {Campbell}}, \bibinfo {author} {\bibfnamefont
  {Y.}~\bibnamefont {Chen}},  \emph {et~al.},\ }\href@noop {} {\bibfield
  {journal} {\bibinfo  {journal} {Nature}\ }\textbf {\bibinfo {volume} {534}},\
  \bibinfo {pages} {222} (\bibinfo {year} {2016})}\BibitemShut {NoStop}%
\bibitem [{\citenamefont {Abadi}\ \emph {et~al.}(2016)\citenamefont {Abadi},
  \citenamefont {Barham}, \citenamefont {Chen}, \citenamefont {Chen},
  \citenamefont {Davis}, \citenamefont {Dean}, \citenamefont {Devin},
  \citenamefont {Ghemawat}, \citenamefont {Irving}, \citenamefont {Isard} \emph
  {et~al.}}]{abadi2016tensorflow}%
  \BibitemOpen
  \bibfield  {author} {\bibinfo {author} {\bibfnamefont {M.}~\bibnamefont
  {Abadi}}, \bibinfo {author} {\bibfnamefont {P.}~\bibnamefont {Barham}},
  \bibinfo {author} {\bibfnamefont {J.}~\bibnamefont {Chen}}, \bibinfo {author}
  {\bibfnamefont {Z.}~\bibnamefont {Chen}}, \bibinfo {author} {\bibfnamefont
  {A.}~\bibnamefont {Davis}}, \bibinfo {author} {\bibfnamefont
  {J.}~\bibnamefont {Dean}}, \bibinfo {author} {\bibfnamefont {M.}~\bibnamefont
  {Devin}}, \bibinfo {author} {\bibfnamefont {S.}~\bibnamefont {Ghemawat}},
  \bibinfo {author} {\bibfnamefont {G.}~\bibnamefont {Irving}}, \bibinfo
  {author} {\bibfnamefont {M.}~\bibnamefont {Isard}},  \emph {et~al.},\ }in\
  \href@noop {} {\emph {\bibinfo {booktitle} {OSDI}}},\ Vol.~\bibinfo {volume}
  {16}\ (\bibinfo {year} {2016})\ pp.\ \bibinfo {pages} {265--283}\BibitemShut
  {NoStop}%
\bibitem [{\citenamefont {Gogate}\ and\ \citenamefont
  {Dechter}(2004)}]{gogate2004complete}%
  \BibitemOpen
  \bibfield  {author} {\bibinfo {author} {\bibfnamefont {V.}~\bibnamefont
  {Gogate}}\ and\ \bibinfo {author} {\bibfnamefont {R.}~\bibnamefont
  {Dechter}},\ }in\ \href@noop {} {\emph {\bibinfo {booktitle} {Proc CUAI}}}\
  (\bibinfo {year} {2004})\ pp.\ \bibinfo {pages} {201--208}\BibitemShut
  {NoStop}%
\bibitem [{Das()}]{Dask17}%
  \BibitemOpen
  \href@noop {} {\enquote {\bibinfo {title} {Dask},}\ }\bibinfo {howpublished}
  {\url{https://dask.pydata.org/en/latest/}},\ \bibinfo {note} {accessed:
  12-13-2017}\BibitemShut {NoStop}%
\bibitem [{\citenamefont {Bishop}\ and\ \citenamefont
  {ligne)}(2006)}]{bishop_pattern_2006}%
  \BibitemOpen
  \bibfield  {author} {\bibinfo {author} {\bibfnamefont {C.}~\bibnamefont
  {Bishop}}\ and\ \bibinfo {author} {\bibfnamefont {S.~S.~e.}\ \bibnamefont
  {ligne)}},\ }\href@noop {} {\emph {\bibinfo {title} {Pattern recognition and
  machine learning}}},\ Vol.~\bibinfo {volume} {4}\ (\bibinfo  {publisher}
  {Springer New York},\ \bibinfo {year} {2006})\BibitemShut {NoStop}%
\bibitem [{\citenamefont {Boixo}\ \emph {et~al.}(2014)\citenamefont {Boixo},
  \citenamefont {R{\o}nnow}, \citenamefont {Isakov}, \citenamefont {Wang},
  \citenamefont {Wecker}, \citenamefont {Lidar}, \citenamefont {Martinis},\
  and\ \citenamefont {Troyer}}]{boixo2014evidence}%
  \BibitemOpen
  \bibfield  {author} {\bibinfo {author} {\bibfnamefont {S.}~\bibnamefont
  {Boixo}}, \bibinfo {author} {\bibfnamefont {T.~F.}\ \bibnamefont
  {R{\o}nnow}}, \bibinfo {author} {\bibfnamefont {S.~V.}\ \bibnamefont
  {Isakov}}, \bibinfo {author} {\bibfnamefont {Z.}~\bibnamefont {Wang}},
  \bibinfo {author} {\bibfnamefont {D.}~\bibnamefont {Wecker}}, \bibinfo
  {author} {\bibfnamefont {D.~A.}\ \bibnamefont {Lidar}}, \bibinfo {author}
  {\bibfnamefont {J.~M.}\ \bibnamefont {Martinis}}, \ and\ \bibinfo {author}
  {\bibfnamefont {M.}~\bibnamefont {Troyer}},\ }\href@noop {} {\bibfield
  {journal} {\bibinfo  {journal} {Nature Physics}\ }\textbf {\bibinfo {volume}
  {10}},\ \bibinfo {pages} {218} (\bibinfo {year} {2014})}\BibitemShut
  {NoStop}%
\bibitem [{\citenamefont {Kask}\ \emph {et~al.}(2001)\citenamefont {Kask},
  \citenamefont {Dechter}, \citenamefont {Larrosa},\ and\ \citenamefont
  {Cozman}}]{kask2001bucket}%
  \BibitemOpen
  \bibfield  {author} {\bibinfo {author} {\bibfnamefont {K.}~\bibnamefont
  {Kask}}, \bibinfo {author} {\bibfnamefont {R.}~\bibnamefont {Dechter}},
  \bibinfo {author} {\bibfnamefont {J.}~\bibnamefont {Larrosa}}, \ and\
  \bibinfo {author} {\bibfnamefont {F.}~\bibnamefont {Cozman}},\ }\href@noop {}
  {\bibfield  {journal} {\bibinfo  {journal} {Artif. Intel}\ }\textbf {\bibinfo
  {volume} {125}},\ \bibinfo {pages} {131} (\bibinfo {year}
  {2001})}\BibitemShut {NoStop}%
\bibitem [{\citenamefont {Pearl}(1982)}]{pearl1982reverend}%
  \BibitemOpen
  \bibfield  {author} {\bibinfo {author} {\bibfnamefont {J.}~\bibnamefont
  {Pearl}},\ }\href@noop {} {\emph {\bibinfo {title} {Reverend Bayes on
  inference engines: A distributed hierarchical approach}}}\ (\bibinfo
  {publisher} {UCLA},\ \bibinfo {year} {1982})\BibitemShut {NoStop}%
\bibitem [{\citenamefont {Bethe}(1935)}]{bethe1935statistical}%
  \BibitemOpen
  \bibfield  {author} {\bibinfo {author} {\bibfnamefont {H.~A.}\ \bibnamefont
  {Bethe}},\ }\href@noop {} {\bibfield  {journal} {\bibinfo  {journal}
  {Proceedings of the Royal Society of London. Series A, Mathematical and
  Physical Sciences}\ }\textbf {\bibinfo {volume} {150}},\ \bibinfo {pages}
  {552} (\bibinfo {year} {1935})}\BibitemShut {NoStop}%
\bibitem [{\citenamefont {M{\'e}zard}\ \emph {et~al.}(1987)\citenamefont
  {M{\'e}zard}, \citenamefont {Parisi},\ and\ \citenamefont
  {Virasoro}}]{mezard1987spin}%
  \BibitemOpen
  \bibfield  {author} {\bibinfo {author} {\bibfnamefont {M.}~\bibnamefont
  {M{\'e}zard}}, \bibinfo {author} {\bibfnamefont {G.}~\bibnamefont {Parisi}},
  \ and\ \bibinfo {author} {\bibfnamefont {M.}~\bibnamefont {Virasoro}},\
  }\href@noop {} {\emph {\bibinfo {title} {Spin glass theory and beyond: An
  Introduction to the Replica Method and Its Applications}}},\ Vol.~\bibinfo
  {volume} {9}\ (\bibinfo  {publisher} {World Scientific Publishing Co Inc},\
  \bibinfo {year} {1987})\BibitemShut {NoStop}%
\bibitem [{\citenamefont {Lidar}(2004)}]{lidar_quantum_2004}%
  \BibitemOpen
  \bibfield  {author} {\bibinfo {author} {\bibfnamefont {D.~A.}\ \bibnamefont
  {Lidar}},\ }\href@noop {} {\bibfield  {journal} {\bibinfo  {journal} {New J.
  Phys}\ }\textbf {\bibinfo {volume} {6}},\ \bibinfo {pages} {167} (\bibinfo
  {year} {2004})}\BibitemShut {NoStop}%
\bibitem [{\citenamefont {Geraci}\ and\ \citenamefont
  {Lidar}(2010)}]{geraci_classical_2010}%
  \BibitemOpen
  \bibfield  {author} {\bibinfo {author} {\bibfnamefont {J.}~\bibnamefont
  {Geraci}}\ and\ \bibinfo {author} {\bibfnamefont {D.~A.}\ \bibnamefont
  {Lidar}},\ }\href {\doibase 10.1088/1367-2630/12/7/075026} {\bibfield
  {journal} {\bibinfo  {journal} {New J. Phys}\ }\textbf {\bibinfo {volume}
  {12}},\ \bibinfo {pages} {075026} (\bibinfo {year} {2010})}\BibitemShut
  {NoStop}%
\bibitem [{\citenamefont {De~las Cuevas}\ \emph {et~al.}(2011)\citenamefont
  {De~las Cuevas}, \citenamefont {Van~den Nest}, \citenamefont {Martin-Delgado}
  \emph {et~al.}}]{de2011quantum}%
  \BibitemOpen
  \bibfield  {author} {\bibinfo {author} {\bibfnamefont {G.}~\bibnamefont
  {De~las Cuevas}}, \bibinfo {author} {\bibfnamefont {M.}~\bibnamefont {Van~den
  Nest}}, \bibinfo {author} {\bibfnamefont {M.}~\bibnamefont {Martin-Delgado}},
   \emph {et~al.},\ }\href@noop {} {\bibfield  {journal} {\bibinfo  {journal}
  {New J. Phys.}\ }\textbf {\bibinfo {volume} {13}},\ \bibinfo {pages} {093021}
  (\bibinfo {year} {2011})}\BibitemShut {NoStop}%
\bibitem [{\citenamefont {Gottesman}(1998)}]{gottesman1998heisenberg}%
  \BibitemOpen
  \bibfield  {author} {\bibinfo {author} {\bibfnamefont {D.}~\bibnamefont
  {Gottesman}},\ }\href@noop {} {\bibfield  {journal} {\bibinfo  {journal}
  {arXiv:quant-ph/9807006}\ } (\bibinfo {year} {1998})}\BibitemShut {NoStop}%
\bibitem [{\citenamefont {Fujii}\ and\ \citenamefont
  {Morimae}(2017)}]{fujii2013quantum}%
  \BibitemOpen
  \bibfield  {author} {\bibinfo {author} {\bibfnamefont {K.}~\bibnamefont
  {Fujii}}\ and\ \bibinfo {author} {\bibfnamefont {T.}~\bibnamefont
  {Morimae}},\ }\href {\doibase 10.1088/1367-2630/aa5fdb} {\bibfield  {journal}
  {\bibinfo  {journal} {New J. Phys.}\ }\textbf {\bibinfo {volume} {19}},\
  \bibinfo {pages} {033003} (\bibinfo {year} {2017})}\BibitemShut {NoStop}%
\bibitem [{\citenamefont {Goldberg}\ and\ \citenamefont
  {Guo}(2014)}]{goldberg_complexity_2014}%
  \BibitemOpen
  \bibfield  {author} {\bibinfo {author} {\bibfnamefont {L.~A.}\ \bibnamefont
  {Goldberg}}\ and\ \bibinfo {author} {\bibfnamefont {H.}~\bibnamefont {Guo}},\
  }\href {http://arxiv.org/abs/1409.5627} {\bibfield  {journal} {\bibinfo
  {journal} {arXiv:1409.5627}\ } (\bibinfo {year} {2014})}\BibitemShut
  {NoStop}%
\end{thebibliography}%

\end{document}